\documentclass[aps,prb,a4paper,twocolumn,10pt,superscriptaddress,notitlepage]{revtex4-2}

\newcommand{\papertitle}{Cryo-Near-Field Photovoltage Microscopy of Heavy-Fermion Twisted Symmetric Trilayer Graphene}

\usepackage[utf8]{inputenc}
\usepackage{graphicx}
\usepackage{wrapfig}
\usepackage{amsmath}
\usepackage{amssymb}
\usepackage{natbib}
\usepackage{float}
\usepackage{tikz}
\usepackage{siunitx}
\usepackage{dcolumn}
\usepackage{lmodern}
\usepackage[T1]{fontenc}
\usepackage[unicode=true,
	    colorlinks=true,
	    linkcolor=blue,
	    citecolor=blue,
	    urlcolor=blue]{hyperref} 
\usepackage{sansmath}
\usepackage{blindtext}
\usepackage[normalem]{ulem}

\usepackage{chemformula}
\usepackage{empheq}
\usepackage{comment}
\usepackage{bm}
\usepackage{gensymb}

\usepackage[misc]{ifsym}
\usepackage{fontawesome}

\usepackage{lineno}

\setcitestyle{super}

\usepackage{xr}
\makeatletter

\newcommand*{\addFileDependency}[1]{
\typeout{(#1)}
\@addtofilelist{#1}
\IfFileExists{#1}{}{\typeout{No file #1.}}
}\makeatother

\newcommand*{\myexternaldocument}[1]{%
\externaldocument{#1}%
\addFileDependency{#1.tex}%
\addFileDependency{#1.aux}%
}

\myexternaldocument{supplement}

\usepackage[top=2.5cm,bottom=2.5cm,left=1.5cm,right=1.5cm]{geometry}

\begin{document}

\title{\Large\textsf{\papertitle}}

\author{Sergi Batlle Porro}
\affiliation{\footnotesize ICFO-Institut de Ciències Fotòniques, The Barcelona Institute of Science and Technology, Av. Carl Friedrich Gauss 3, 08860 Castelldefels (Barcelona),~Spain}

\author{Dumitru C\u{a}lug\u{a}ru}
\affiliation{\footnotesize Department of Physics, Princeton University, Princeton, NJ 08544,~USA}

\author{Haoyu Hu}
\affiliation{\footnotesize Donostia International Physics Center (DIPC),
Paseo Manuel de Lardizàbal. 20018, San Sebastiàn,~Spain}

\author{Roshan Krishna Kumar}
\affiliation{\footnotesize ICFO-Institut de Ciències Fotòniques, The Barcelona Institute of Science and Technology, Av. Carl Friedrich Gauss 3, 08860 Castelldefels (Barcelona),~Spain}

\author{Niels C.H. Hesp}
\affiliation{\footnotesize ICFO-Institut de Ciències Fotòniques, The Barcelona Institute of Science and Technology, Av. Carl Friedrich Gauss 3, 08860 Castelldefels (Barcelona),~Spain}

\author{Kenji Watanabe}
\affiliation{\footnotesize Research Center for Functional Materials, National Institute for Materials Science, 1-1 Namiki, Tsukuba 305-0044,~Japan}

\author{Takashi Taniguchi}
\affiliation{\footnotesize International Center for Materials Nanoarchitectonics, National Institute for Materials Science,  1-1 Namiki, Tsukuba 305-0044,~Japan}

\author{B. Andrei Bernevig}
\affiliation{\footnotesize Department of Physics, Princeton University, Princeton, NJ 08544,~USA}
\affiliation{\footnotesize Donostia International Physics Center (DIPC),
Paseo Manuel de Lardizàbal. 20018, San Sebastiàn,~Spain}
\affiliation{\footnotesize IKERBASQUE, Basque Foundation for Science, 48013 Bilbao,~Spain}

\author{Petr Stepanov}
\email{pstepano@nd.edu}
\affiliation{\footnotesize ICFO-Institut de Ciències Fotòniques, The Barcelona Institute of Science and Technology, Av. Carl Friedrich Gauss 3, 08860 Castelldefels (Barcelona),~Spain}
\affiliation{\footnotesize Department of Physics and Astronomy, University of Notre Dame, Notre Dame, IN 46556,~USA}
\affiliation{\footnotesize Stavropoulos Center for Complex Quantum Matter, University of Notre Dame, Notre Dame, IN 46556,~USA}

\author{Frank H.L. Koppens}
\email{frank.koppens@icfo.eu}
\affiliation{\footnotesize ICFO-Institut de Ciències Fotòniques, The Barcelona Institute of Science and Technology, Av. Carl Friedrich Gauss 3, 08860 Castelldefels (Barcelona),~Spain}
\affiliation{\footnotesize ICREA-Institució Catalana de Recerca i Estudis Avançats, 08010 Barcelona,~Spain}



\begin{abstract}
\vspace*{0.2cm}
Ever since the initial experimental observation of correlated insulators and superconductivity in the flat Dirac bands of magic angle twisted bilayer graphene, a search for the microscopic description that explains its strong electronic interactions has begun. While the seemingly disagreeing electronic transport and scanning tunneling microscopy experiments suggest a dichotomy between local and extended electronic orbitals, definitive experimental evidence merging the two patterns together has been much sought after. Here, we report on the local photothermoelectric measurements in the flat electronic bands of twisted symmetric trilayer graphene (TSTG). We use a cryogenic scanning near-field optical microscope with an oscillating atomic force microscopy (AFM) tip irradiated by the infrared photons to create a nanoscopic hot spot in the planar samples, which generates a photocurrent that we probe globally. We observe a breakdown of the non-interacting Mott formalism at low temperatures ($\sim$10 K), signaling the importance of the electronic interactions. Our measurements reveal an overall negative offset of the Seebeck coefficient and significant peaks of the local photovoltage values at all positive integer fillings of the TSTG's moiré superlattice, further indicating a substantial deviation from the classical two-band semiconductor Seebeck response. We explain these observations using the interacting topological heavy-fermion model. In addition, our data reveal a spatial variation of the relative interaction strength dependent on the measured local twist angle (1.2$^\circ$ - 1.6$^\circ$). Our findings provide experimental evidence of heavy fermion behaviour in the topological flat bands of moiré graphene and epitomize an avenue to apply local thermoelectric measurements to other strongly correlated materials in the disorder-free limit.
\end{abstract}

\maketitle

\noindent
\textbf{Introduction.} Over the past decades, heavy-fermion compounds have attracted significant attention \cite{wirth_exploring_2016, stewart_heavy-fermion_1984}. Early experiments demonstrated \textit{f}-electron virtual bound states in CeAl$_3$ that obey the Landau theory of Fermi liquids\cite{landau_lev_davidovich_theory_1956} with the strong enhancement of the charge carriers' effective mass - heavy fermions \cite{andres_4_1975}. By virtue of that, these systems exhibit giant anomalies, including long-range magnetic orders and unconventional superconductivity\cite{steglich_superconductivity_1979, radovan_magnetic_2003}. While offering a fascinating playground for experimentalists, heavy-fermion compounds have posed significant challenges in achieving complete theoretical descriptions of their exotic order parameters.

\begin{figure*}[t]
    \centering
    \includegraphics[scale=0.58]{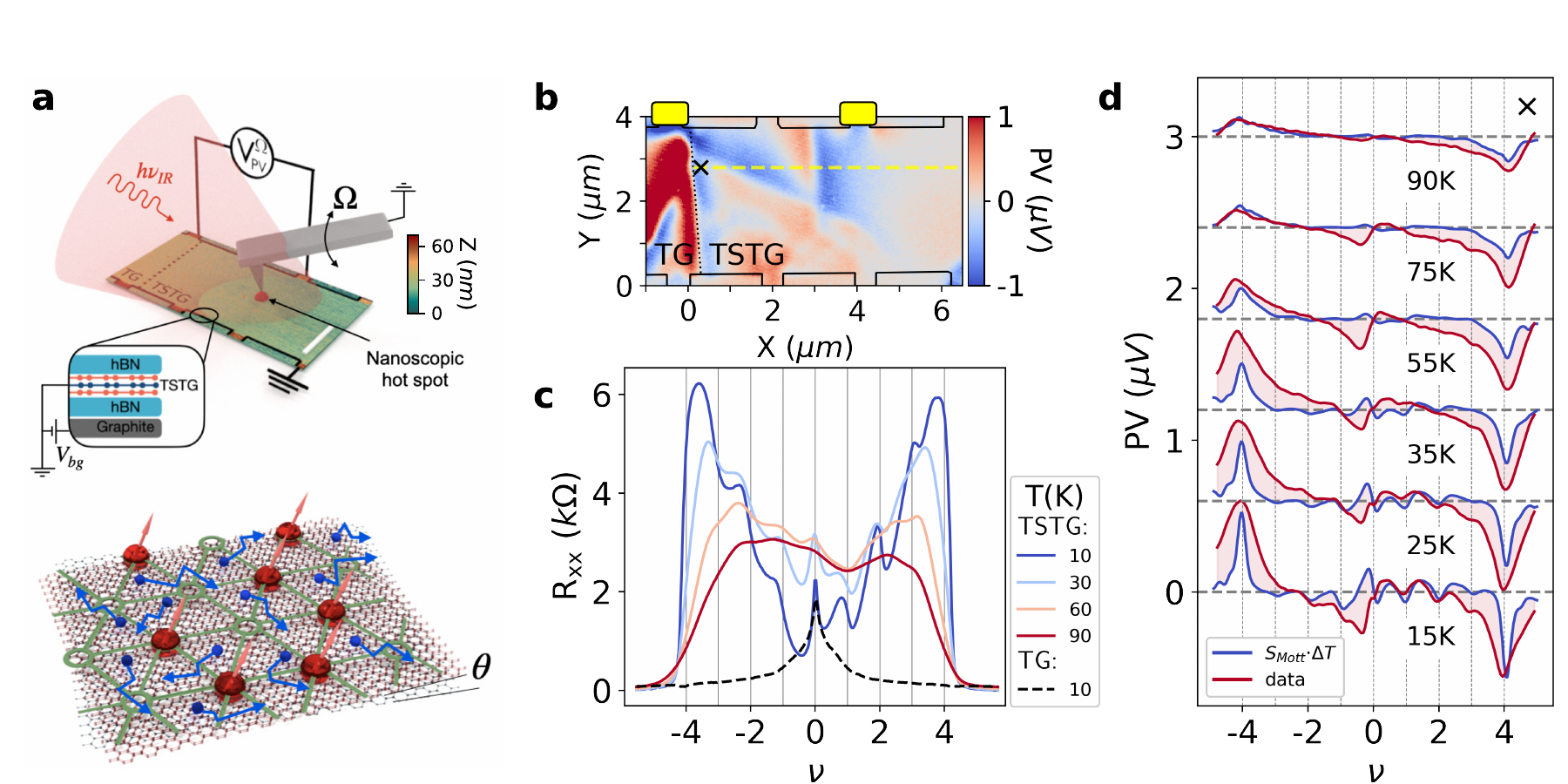}
    \caption{
    	\textbf{Experimental setup and Mott formula breakdown.} \textbf{a} (Top) Schematics of the photovoltage nanoscopy experiment. The color plane shows the AFM topography height of the sample, as indicated by the colorbar. Focused infrared light irradiates a metallic AFM tip. $V^{\Omega}_{PV}$ is measured on the global probes using a lock-in technique at the frequency of the tip oscillations $\Omega$. Scale bar is 2 $\mu$m. (Bottom) Schematic representation of topological heavy fermion model. Transparent red spheres represent short-lived localized moments of $\textit{f}$ electrons. The small blue spheres represent long-lived itinerant electrons. The green circles denote the AA sites of the superlattice, while the green lines show AB/BA domain boundaries. \textbf{b} $PV$ as a function of the tip position taken at $\nu$ = +1 using a pair of contacts highlighted by the yellow boxes. The black dashed line in the left part of the scan indicates a boundary between TSTG faulted at the magic angle and trilayer graphene (TG) of an unknown twist angle. Note the large positive/negative $PV$ signal on the left/right side of the interface, respectively. \textbf{c} $R_{xx}$ versus band filling $\nu$ taken at different temperatures within the TSTG region (solid lines) and within the TG region (dashed line). \textbf{d} Local $PV$ vs. band filling at different temperatures (red lines) taken at the tip position indicated by a cross in \textbf{b}. Blue lines indicate $S_{\mathrm{Mott}}$ in the case of classical metallic electron diffusion computed via Mott formula\cite{jonson_mott_1980} from the resistivity data shown in \textbf{c} (see SI Section III).}\label{Mott breakdown}  
\end{figure*}

The emerging field of twistronics has introduced a novel approach to exploring strongly correlated physics within two-dimensional planes of twisted or closely aligned van der Waals layers\cite{bistritzer_moire_2011, andrei_graphene_2020, balents_superconductivity_2020}. In particular, some of the moiré superlattices host superconductivity\cite{cao_unconventional_2018, yankowitz_tuning_2019, lu_superconductors_2019, po_origin_2018, dodaro_phases_2018, xie_nature_2020, oh_evidence_2021}, correlated insulators\cite{cao_correlated_2018, nuckolls_strongly_2020}, nontrivial topology\cite{song_topology_2019, wagner_global_2022, lian_twisted_2021, wu_chern_2021} and orbital ferromagnetism\cite{serlin_intrinsic_2020, sharpe_emergent_2019, polshyn_electrical_2020}, all originating from their flat bands that emerge around charge neutrality. Semiconducting moiré crystals based on transition metal dichalcogenides have been constructed to probe the limits of the Hubbard model\cite{tang_simulation_2020} and even to create a synthetic Kondo lattice \cite{zhao_gate-tunable_2023}. Magic angle twisted symmetric trilayer graphene (TSTG) has emerged as a condensed matter system that exhibits signatures of the BEC-BCS transition\cite{park_tunable_2021} as well as signatures of the spin-triplet superconductivity\cite{cao_pauli-limit_2021}, albeit leaving open questions about the nature of its ground states. Spectroscopic evidence for the local Wannier orbitals in these systems was delivered by scanning tunneling microscopy experiments \cite{jiang_charge_2019, kerelsky_maximized_2019, wong_cascade_2020}. While electronic transport experiments suggest delocalized electrons form strongly correlated phases\cite{cao_unconventional_2018, saito_isospin_2021}, local probe spectroscopy points to the narrowly confined charge carriers around the AA sites of the graphene moiré superlattice\cite{kerelsky_maximized_2019, zondiner_cascade_2020,rozen_entropic_2021, choi_correlation-driven_2021}.

\begin{figure*}[t]
    \centering
    \includegraphics[scale=0.58]{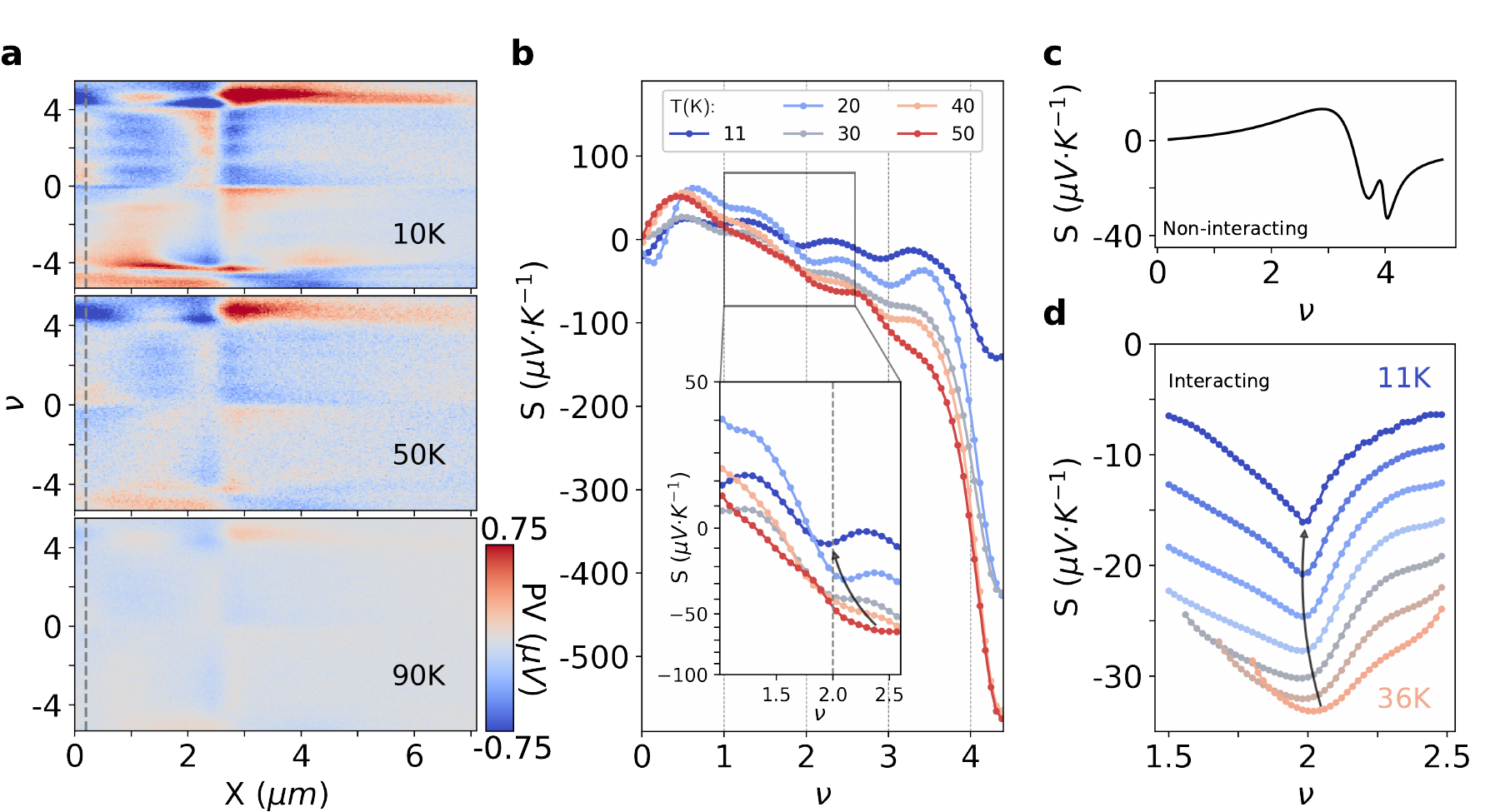}
    \caption{
    	\textbf{The Seebeck effect of TSTG.} \textbf{a} Contourplots of $PV$ vs. $\nu$ and the tip position $x$ along the yellow dashed line in Fig.~\ref{Mott breakdown}b taken at 10, 50 and 90 K. Note a general gain of the $PV$ magnitude as well as the formation of the horizontal stripe-like features at integer fillings when $T$ decreases to 10 K. \textbf{b} $S$ versus $\nu$ along the dashed line in \textbf{a}. Here we observe a negative vertical offset of $S$ and an emergence of a pronounced sawtooth-like oscillations of $S$ at low temperatures. (Inset) A zoom-in semi-logarithmic $S$ vs. $\nu$ plot. The black arrow traces a local minimum of $S$ at $\nu=+2$ for all five temperature points. \textbf{c} Theoretical calculations of the Seebeck coefficient based on the THF model for TSTG\cite{calugaru_seebeck} without interactions at $T=15$ K. \textbf{d} With interactions at $\nu$ = +2 at $T$=11.2, 15.4, 19.6, 23.8, 28.0, 32.2, 36.4 K. The black arrow indicate local minima in $S$. Note a close resemblance to the experimental minima in the inset of \textbf{b}.}\label{Comparison}
\end{figure*}

Here, we report on the photothermoeletric nanoscopy of TSTG, which probes the Seebeck coefficient ($S$) locally by a nanometer-size hot spot ($\sim$20 nm). By measuring local photovoltage ($PV\propto S$), our data indicate the coexistence of charge-one excitations with markedly different lifetimes, namely heavy holes and light electrons, for positive fillings of the moiré graphene minibands. We observe a substantial deviation of the measured $PV$ from the classical electron diffusion model\cite{jonson_mott_1980}. In particular, within the magic angle domains, the measured $PV$ exhibits minima at integer fillings $\nu$ = +1, +2, and +3, and features an overall negative Seebeck coefficient, indicating that electrons (as opposed to holes) dominate thermoelectric transport at all positive fillings. 

To explain this observation, we compare our experimental results with the recently proposed topological heavy fermion (THF) model for moiré graphene\cite{song_magic-angle_2022, yu_magic-angle_2023, calugaru_seebeck}. The cohesion between the experimental observations and the theoretical model is consistent with a heavy fermion description of TSTG. By probing TSTG's Fermi surface with Hall transport we find a significant decrease of the Hall charge carrier density at the integer fillings, which which points at the localization of the heavy electrons on the topological flat \textit{f}-bands and the inception of the correlated insulator behavior within the TSTG flat bands. Furthermore, the spatially resolved measurements allow us to associate the unconventional $PV$ response with the local twist angle, identifying a range of angles for which heavy fermion features are noticeable. Taken together, our observations establish TSTG near the magic angle as a highly tuneable heavy fermion condensed matter system and the near-field $PV$ nanoscopy as a sensitive local probe for the strength of the electronic interactions.

\textbf{Local photothermoeletric nanoscopy.} We use a commercial cryogenic scanning near-field optical microscope (cryo-neasSCOPE) with an oscillating metallic AFM tip in the tapping regime to create a local hot-spot in planar samples (Fig.~\ref{Mott breakdown}a). Our devices are fabricated using a standard dry transfer "cut-and-stack" technique (see Methods) and consist of hBN-encapsulated TSTG graphene with a global graphite gate (the thicknesses of hBN layers are 20 nm for the top and 17 nm for the bottom layers, respectively). Through the heating of electrons, we achieve thermoelectric transport in the inhomogeneous sample of TSTG and generate a finite voltage drop across the globally positioned electrical contacts to our sample ($V_{PV}^\Omega$ in Fig.~\ref{Mott breakdown}a)\cite{woessner_near-field_2016}. The sample studied in this experiment consists of two parts. The right half of the sample is twisted close to the magic angle condition. In contrast, the left side of the device is faulted at an angle far away from the magic angle condition due to a possible microfolding of the middle layer during the stacking process (dashed line in AFM image of Fig.~\ref{Mott breakdown}a denotes the interface). This is further elucidated by the electron transport measurements shown in Fig.~\ref{Mott breakdown}c. In the TSTG part, electronic transport shows the formation of the resistance peaks close to all integer $\nu$ signaling the onset of the correlated insulating behavior\cite{kim_evidence_2023}. For the remainder of our discussion, we will focus exclusively on the magic angle TSTG part of our device.

The red curves in Fig.~\ref{Mott breakdown}d show the temperature dependence of $V_{PV}^{\Omega} \equiv PV\propto{S}\Delta T$ as a function of the band filling $\nu$ taken at a point shown by the cross in Fig.~\ref{Mott breakdown}b (see Methods and SI Section V for more detailed information about the conversion of the backgate voltage into the band filling factor $\nu$). Here, we anticipate that the locally enhanced electromagnetic field leads to a local increase of the electron temperature in the TSTG layer $\Delta T$ ($\Delta T \propto 1/T$, see SI Section II for more information). First, we note that the experimentally obtained $PV$ exhibits strong negative (positive) values for the electron (hole)-doped flat bands, respectively. The negative values are steadily growing in magnitude with decreasing $T$ and increasing doping (until the full fillings) signaling electron dominated transport for positive $\nu$. Around integer fillings, the flat bands of TSTG develop correlated phases at low $T$, while the Dirac cone is approximately pinned in the gap due to its high Fermi velocity. Conventionally, this would result in a Seebeck coefficient that changes sign at each positive integer filling from positive to negative as $\nu$ is increased. We observe an opposite trend. At $\nu \gtrsim +1.5$, $PV$ significantly deviates towards overall negative values. 

Second, we compare these measurements with the classical electron diffusion model (blue lines) described by the Mott formalism $S_{\mathrm{Mott}}\sim T \times d(\log(\sigma_{xx}))/dV_{bg}$ (see SI Section II for more information), where $S_{\mathrm{Mott}}$ is the Seebeck coefficient, $\sigma_{xx}$ is the longitudinal conductivity and $V_{bg}$ is the backgate voltage. At 90 K, the curves match reasonably well, confirming the photothermoelectric effect mechanism and signaling an increased role of the TSTG's Dirac cone, which consistently favors a negative Seebeck coefficient\cite{calugaru_seebeck}. Upon decreasing $T$, however, we find a considerable deviation of the red curves from the Mott formula for all temperatures when $\nu \gtrsim +2$ (shaded regions in  Fig.~\ref{Mott breakdown}d highlight the difference). We note that $S_{\mathrm{Mott}}$ generally follows a conventional trend, switching from positive to negative values across every positive integer fillings. As a consequence of its macroscopic nature, the Mott formula does not provide any details on the microscopic mechanisms governing thermopower enhancement. However, its breakdown\cite{ghawri2022breakdown}, along with a significant negative offset of the Seebeck values in TSTG's moiré bands, points to a strongly unconventional thermoelectric transport mechanism dominating at low temperatures.

\begin{figure*}[t]
    \centering
    \includegraphics[scale=0.88]{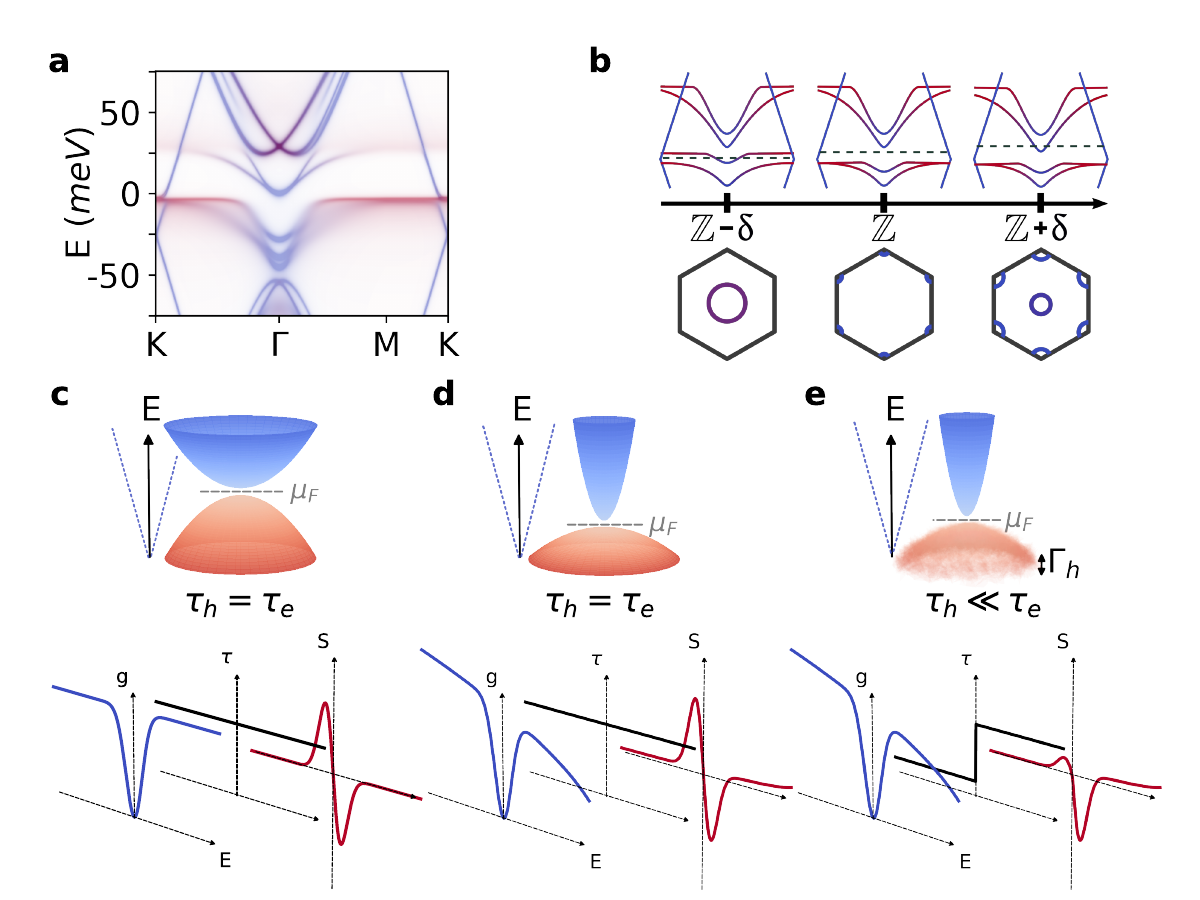}
    \caption{
    	\textbf{Bands, Fermi surface and the carrier's lifetime in the Seebeck effect.} \textbf{a} \textbf{k}-resolved spectral function (color) $A(\omega, \textbf{k})$ for $\nu = 1.8$ at $T$=11 K. Here, the blue color represents the long-lived $c$ ($d$)-bands, while the red color corresponds to the short-lived $f$-bands, respectively. \textbf{b} (\textbf{top}) Schematics showing the evolution of the flatbands with doping around an integer filling. (\textbf{bottom}) Schematics of the Fermi surface in the moiré Brillouin zone. \textbf{c (top)} A symmetric density of states $g(E)$ resulting from the two parabolic bands shown in the inset for a small gap where the band lifetimes below ($E<\mu_F$) and above ($E>\mu_F$) the Fermi level are equal ($\tau_{h} = \tau_{e}$). Dashed blue lines represent a section of the TSTG Dirac cone dispersion. \textbf{(bottom)} The resulting Seebeck coefficient values vs. carrier energy $E$. \textbf{d} Same as \textbf{c} but with an asymmetric density of states resulting from different effective mass of the bands. Note the similarity of the Seebeck coefficient with the one in \textbf{c}.  \textbf{e} Same as \textbf{d} but here $\tau_{h} \ll \tau_{e}$ (\textbf{top right}). The carrier's lifetime for $E<\mu_F$ is drastically smaller than for $E>\mu_F$. \textbf{(bottom)}. Resulting Seebeck coefficient vs. energy $E$. Note the significant asymmetry in the Seebeck coefficient's absolute values for $E<\mu_F$ and $E>\mu_F$.}\label{Lifetime}
\end{figure*}

To elucidate the unconventional Seebeck coefficient, we perform spatially resolved and temperature-dependent studies of TSTG. We examine $PV$ versus $\nu$ and the tip position $x$ along the yellow dashed line shown in Fig.~\ref{Mott breakdown}b. A set of three maps is presented in Fig.~\ref{Comparison}a at temperatures of 10, 50, and 90 K. Serving as a measure of the Seebeck coefficient, $PV$ reveals rather unexpected patterns. With decreasing temperature, we observe an enhancement of $PV$ accompanied by the formation of spatially resolved sawtooth-like minima at integer fillings $\nu$= +1, +2, and +3, spanning horizontally across these maps. Close to the interface ($x=0~\mu$m), the Seebeck effect from the TG part of our sample influences the thermoelectric transport in the TSTG part. Around the charge neutrality point, we detect a weak positive $PV$ at small positive $\nu$. However, farther from the TSTG/TG interface and before the TSTG grain boundary (i. e. at $x \approx 0.5 - 2~ \mathrm{\mu m}$), we observe an all-negative $PV$ at positive $\nu$ inside the electron-like flat band.

We further explore the minima in Fig.~\ref{Comparison}b by plotting the linecuts of the measured $S$ vs. $\nu$ at different temperatures along the vertical dashed line in Fig.~\ref{Comparison}a (here, $S=PV/\Delta T$). First, in contrast to the classical Seebeck response of a semimetal reported in the single-layer graphene\cite{woessner_near-field_2016, gabor_hot_2011, ma_tuning_2016}, we observe an overall negative vertical offset of $S$ for the electron-doped flat band. The trend of increasing Seebeck magnitude and overall negative values in the electron-doped region strongly points to the electron-dominated thermoelectric transport across the entire electron-like flat band. Second, a sawtooth-like structure of $S$ at the integer fillings arises due to the opening of correlation induced gaps in the flat bands of TSTG. In the case of a classical two-band semiconductor\cite{dresselhaus2007thermoelectricity}, the Seebeck coefficient vanishes when the Fermi level is tuned across the gap, as thermoelectric transport turns from hole dominated (with a positive $S$) to electron dominated (with a negative $S$). Our observations of the negative offset in $S$ and the sawtooth-like features suggest a pronounced asymmetry between the contribution of electron and hole excitations to thermoelectric transport.\cite{paul_interaction-driven_2022}. 

Following this assumption, the recently introduced topological heavy fermion (THF) model\cite{song_magic-angle_2022, yu_TSTG_THF_2023, datta_heavy_2023, calugaru_seebeck} becomes a possible candidate to explain our experimental observations. At first, we consider the THF model in the absence of interactions at 15 K\cite{novelli_optical_2020, calugaru_seebeck} (Fig.~\ref{Comparison}c). This configuration of the THF model suggests positive Seebeck values for $\nu \approx$ 0 - 3.5, while the experimental Seebeck coefficient in Fig.~\ref{Comparison}b acquires all negative values already at $\nu \gtrsim +1$. Fig.~\ref{Comparison}d shows the theoretically computed Seebeck coefficient of TSTG around the $\nu=+2$ Kramers intervalley coherent (K-IVC) symmetry-broken ground state candidate. The topological heavy fermion model of TSTG naturally consists of three types of carriers: strongly-correlated ($f$) electrons,  weakly-correlated and strongly-dispersive conduction ($c$) and Dirac ($d$) electrons\cite{yu_TSTG_THF_2023}. Around integer fillings, the Dirac cone formed by the $d$-electrons is pinned approximately to the Fermi level, meaning that it gives almost no net contribution to the Seebeck coefficient. In contrast, the correlated $f$ electrons form flat hole bands, while the itinerant $c$ electrons form dispersive electron bands. The diminished lifetime of the $f$-electrons arising from correlation effects naturally reduces their Seebeck effect contribution and leads to an overall electron dominated thermoelectric transport. This consequently leads to a negative Seebeck coefficient for a large doping range around $\nu = +2$ (similar effect also arises around other positive integer fillings $\nu = +1$ and $\nu = +3$), consistent with our observations. Here, similar to the experiment, Seebeck minima shift towards the charge neutrality, as shown by the black arrow in the inset of Fig.~\ref{Comparison}b. The negative peak in the Seebeck coefficient emerges when the chemical potential is at the edge of the electron band. At this point, the broadened hole-band spectral weight "leaks" above the chemical potential, resulting in the system's total filling being slightly smaller than an integer\cite{calugaru_seebeck}.

To further elaborate on the band structure and correlate it with the thermoelectric effect discussed above we show the theoretically obtained spectral function $A(\omega, \textbf{k})$ around $\nu = 2$ in Fig.~\ref{Lifetime}a. The $c$($d$)-bands (blue) form long-lived coherent excitations, while the $f$-bands (red) form short-lived incoherent carriers. When the chemical potential is tuned slightly below the integer fillings ($\nu$ = Z-$\delta$ in Fig.~\ref{Lifetime}b), the Fermi surface is large and charge carriers attain more $f$-character, characterized by short lifetime and decreased band's group velocity. Upon crossing the gap (at $\nu$ = Z+$\delta$), the Fermi surface significantly contracts, and the available charge carriers acquire a $c$ ($d$)-character (see more calculations of the THF band structure in SI Section IV). 

In a simplified way, we elaborate further on the effect of the carrier lifetime $\tau$ in Fig.~\ref{Lifetime}c-e based on a phenomenological two-band model with a small gap. Fig.~\ref{Lifetime}c provides an expected Seebeck coefficient in the case when the density of states $g(E)$ and the lifetime $\tau$ are symmetric ($\tau_h = \tau_e$) around the Fermi level $\mu_F$ (here, $\mu_F$ matches a positive integer filling $\nu=+1, +2, +3$ of the moiré flat band). Using the Kubo formula (see SI Section III), we calculate the corresponding Seebeck coefficient, which is odd around $E=\mu_F$. The Dirac fermion band of TSTG (blue dashed lines in Fig.~\ref{Lifetime}c-e) is approximately pinned at charge neutrality and makes a negligible contribution to the Seebeck coefficient\cite{calugaru_seebeck}. Similar to the case in Fig.~\ref{Lifetime}c, the Seebeck coefficient remains odd when we introduce a small asymmetry in $g(E)$, but keep the lifetime unchanged (Fig.~\ref{Lifetime}d). In contrast, Fig.~\ref{Lifetime}e assumes that the lifetimes $\tau(E)$ across $E=\mu_F$ are no longer equal ($\tau_h \neq \tau_e$). Here, we generally assume that the red band corresponds to the $f$-electrons and the blue band corresponds to the $c$-electrons in Fig.~\ref{Lifetime}a. As a result, the Seebeck coefficient exhibits an enhanced asymmetry with almost vanishing values for holes ($E<\mu_F$) due to the convolution between $g(E)$ and the electron diffusion coefficient $D(E)$, which is linearly proportional to the quasiparticle's lifetime $\tau(E)$. Hence, a major consequence of the lifetime decrease for holes with respect to electrons across $E=\mu_F$ is a significant reduction of the hole-like contribution to the thermoelectric transport for $E<\mu_F$\cite{calugaru_seebeck}.

Consistent with the predictions in Fig.~\ref{Lifetime}c, the measured $S$ acquires overall negative values for $T>20$ K and band fillings $\nu \gtrsim +1$. Furthermore, at the lowest measured temperature $T=11~K$, we observe $S$ offsetting slightly towards the positive values while maintaining the overall negative slope and developing the strongest sawtooth  oscillations. At the lowest temperature, the Seebeck coefficient is expected to experience more of the hole-like contribution as the interaction-induced \textit{f}-electron broadening becomes less significant\cite{calugaru_seebeck} . 

\begin{figure}[!ht]
    \centering
    \includegraphics[scale=0.7]{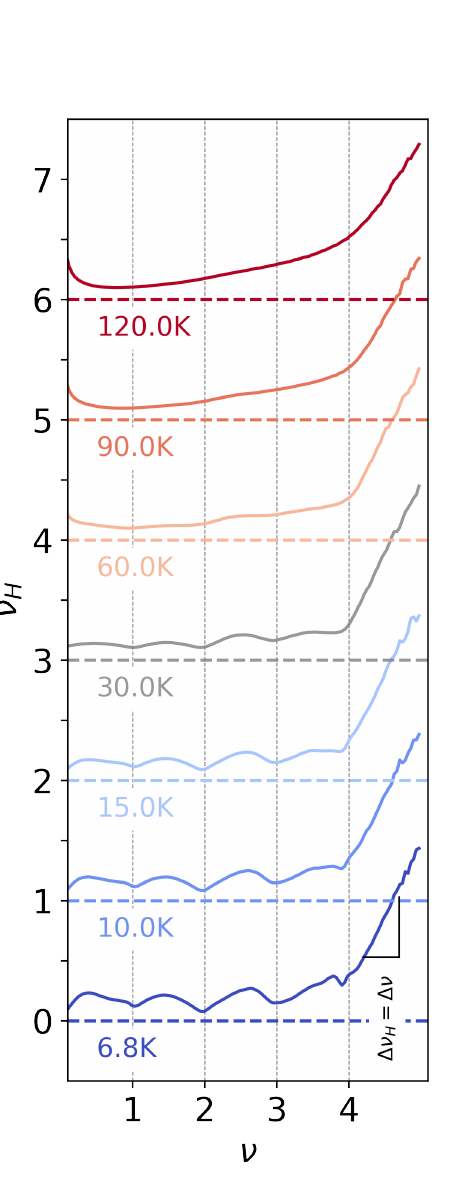}
    \caption{
    	\textbf{TSTG Hall density measurements.} Hall density $\nu_H$ as a function of the free charge carrier density $\nu$ taken at different temperatures between 6.8 and 120 K at $B$=0.5 T. At 6.8 K, Note a significant reduction at $\nu=+1, +2$, and +3 (below $10 \%$ of the free carrier density). See more details in SI Section IX. Curves are shifted vertically by 1 for clarity.}
     \label{Hall}
\end{figure}

\textbf{TSTG Fermi surface.} To further understand the heavy fermion physics in TSTG's flat bands, we perform Hall density measurements in small magnetic fields. In a heuristic heavy-fermion picture, we anticipate that the density of the available charge carriers does not adhere to the relation $\Delta\nu_H$ = $\Delta\nu$ at any $\nu$ inside the TSTG's active bands. While we expect that \textit{f}-electrons localize at AA sites of the superlattice and generally follow $\nu_f$ = +1, +2, and +3 upon crossing the positive integer fillings, the itinerant electron density $\nu_i$ should exhibit a resetting pattern\cite{hu_kondo_2023}. Therefore, it is assumed that the Hall density measurements are sensitive to the presence of the itinerant electrons. Fig.~\ref{Hall}a shows Hall density measurements as a function of the free charge carrier density $\nu$ in the absence of light. At the lowest temperature of 6.8 K, we observe a reduction of the Hall density exactly at the integer fillings albeit it never reaches zero. This observation confirms our expectation of the resetting $\nu_i$. We further corroborate this by examining the temperature dependence of the Hall density. At 60 K, the minima in Hall density at integer fillings completely vanish, suggesting that $f$-band localization no longer occurs due to the enhanced effects of thermal fluctuations. At 120 K, the Hall density shows a smooth dependence on $\nu$, while it does not yet adhere to the relationship $\Delta\nu_H$ = $\Delta\nu$ inside the flat band density band regions. This may indicate residual effects of effective mass enhancement due to band renormalization.

How do our Hall measurements correspond with the local Seebeck coefficient data? The Hall measurements suggest a localization of the $f$-bands, leading to a significant lifetime asymmetry across the positive integer fillings of the bands. This, for example, contributes to a shift of the negative Seebeck peak and the inflection point towards the charge neutrality\cite{calugaru_seebeck} (see the black arrow in Fig.~\ref{Comparison}b). The broadening of the $f$-bands also facilitates the negative offset of the Seebeck coefficient oscillations at the lowest temperatures as it implies a shortened lifetime of the hole excitations, thereby rendering them unavailable for thermoelectric transport. This in turn results in an electron-dominated thermoelectric transport with a negative $S$ at positive integer fillings. On top of the negative offset, the strong $f$-electron interaction effects lead to oscillations in $S$ due to the emergence of the correlated phases around integer fillings in the TSTG flat bands. Consequently, the amplitude of the sawtooth oscillations $PV$  vs. $\nu$ can be considered as a measure of the interaction strength responsible for the emergence of the strongly correlated phases in TSTG. Next, we experimentally explore this measure with the aim of correlating it with the twist angle.

\vspace*{0.2cm}

\begin{figure*}[t]
    \centering
    \includegraphics[scale=0.7]{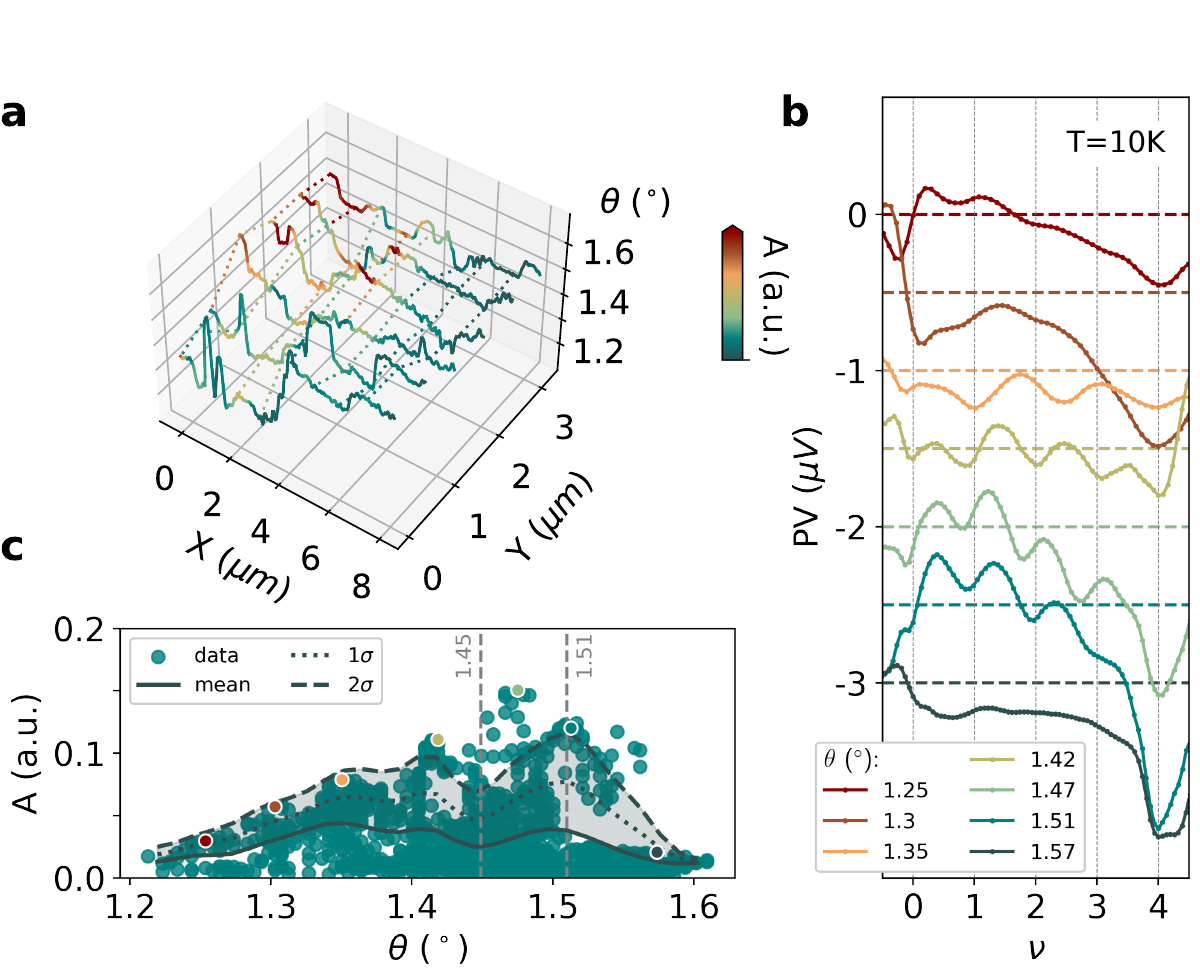}
    \caption{
    	\textbf{Twist angle and correlations.} \textbf{a} 3D wireframe plot of the extracted local twist angles $\theta_{\mathrm{local}}$ (vertical axis). The color represents an amplitude $A$ of the $PV$ vs. $\nu$ oscillations. See SI Section VII for the spatial contourplots used to obtain these data. \textbf{b} $PV$ vs. $\nu$ for several representative $\theta_{\mathrm{local}}$.  \textbf{c} Statistics $A$ vs. $\theta$ for the electron-doped flat band. We observe a clear increase in the magnitude of $A$ in the regions of the sample that are close to the magic angle condition.}\label{Correlations}
\end{figure*}

\textbf{Spatially resolved thermoelectric transport}.
Our real-space local probe allows us to extend thermoelectric measurements into the bulk of the sample. By tracing the position of the $PV$ peaks that correspond to the filling factors $\nu$=$\pm$4, we can extract local twist angles via $n_s = 8 \theta ^2/3a^2$, where $n_s$ is the charge carrier density at $\nu$=$\pm$4 (see Methods and SI Secion V). A representative map of twist angle analysis is shown in Fig.~\ref{Correlations}a. We find a variety of local twist angles in our sample ranging from $\theta$ = 1.2$^\circ$ to 1.6$^\circ$.

Interestingly, we find a clear correlation between the presence of the $PV$ vs. $\nu$ peaks at the integer fillings and the local twist angles (Fig.~\ref{Correlations}b). Specifically, in the areas where the twist angle is significantly below the trilayer magic angle condition (e. g. at $\theta_{\mathrm{local}}$ $\approx$ 1.2$^\circ$ - 1.3$^\circ$), the $PV$ peaks are suppressed signaling the absence of strong correlations occurring near the integer fillings. Furthermore, our data match qualitatively with the high-temperature regime shown in Fig.~\ref{Mott breakdown}d at 90 K. For angles $\theta_{\mathrm{local}}$ $\approx$ 1.35$^\circ$ - 1.51$^\circ$ (Fig.~\ref{Correlations}b), we find  a negative Seebeck coefficient, along with the sawtooth oscillations, pointing to a significant particle-hole lifetime asymmetry between the minibands just below and just above the Fermi level. A more flat and featureless response of $PV$ (e. g. at $\theta_{\mathrm{local}}=$1.25$^\circ$ and 1.57$^\circ$ indicates an absence of the correlated phases at the integer fillings of the band(see also SI Section VII for more spatially resolved data). The locality of our method may also bring some unexpected results in the sign of the measured Seebeck coefficient. In particular, close to the interfaces (e. g. moiré lattice domain walls) we occasionally observe positive $PV$ close to the charge neutrality. However, far from the interfaces similar to Fig.~\ref{Comparison}a, the Seebeck coefficient acquires the expected all-negative values (see SI Section VII for more data). In addition, some regions have $\theta_{\mathrm{local}}$ farther from the magic angle and give a more conventional contribution $S$.

Considering a variety of twist angles in our sample, we map the dependence of the sawtooth amplitude ($A$ in Fig.~\ref{Correlations}a and c) as a function of the twist angle $\theta$. Here, $A$ is extracted as the Fourier amplitude coefficient of oscillations of period $\Delta \nu = 1$, normalized by the total amplitude of the signal at $\nu =4$ (details in Methods and SI section V). This $A$ serves only to represent qualitatively (and not quantitatively) the magnitude of cascade features due to electronic correlations. We observe a maximum of $A$ vs. $\nu$ around the twist angles close to the magic angle condition $\theta_{\mathrm{local}}$ $\approx$ 1.45$^\circ$ - 1.55$^\circ$. This distribution, however, is quite broad pointing at the enhancement of the strong interactions over a large range of the twist angles. Our observation is consistent with previous electronic transport reports which have identified superconductivity in TSTG for $\theta_{\mathrm{local}}$ $\approx$ 1.30$^\circ$ - 1.55$^\circ$\cite{park_tunable_2021, lin_zero-field_2022}. Furthermore, our data qualitatively agrees with the twice the broadening ($2 \Theta_{MB}$) of the magic angle with $\Theta_{MB}=0.15^{\circ}$, denoted as the “magic range"\cite{Bennett_2023}. Here, we emphasize that although $A$ statistically peaks around $1.50^\circ$ for the entire collection of $\theta_{\mathrm{local}}$ in our device and remains consistent with the overall trend for the tip positions in $x \in (0, 4)~\mathrm{\mu m}$, we also find regions that exhibit less correlations while the twist angles are still close to the magic angle value. The reason might be associated with the complex domain structure of the moiré lattice or the difference in the twist angle between the bottom and top graphene layers in TSTG as well as the presence of heterostrain potentials. 

What are the implications of our measurements and their interpretation? Temperature dependencies of $S$ and Hall density measurements firmly suggest an importance of strong electronic interactions. Upon decreasing the temperature, an increasing role of the onsite Coulomb repulsion develops in the minibands with the strong \textit{f}-character. Our data suggests the presence of both localized and itinerant electrons in TSTG around integer fillings. The strong correlation effects experienced by the $f$-electrons lead to an electron-dominated thermoelectric transport at positive integer fillings, and hence to a large negative Seebeck coefficient.

\textbf{Outlook}. Thermoelectric power enhancement comprises a common signature of high-$T_c$ superconductors and other strongly-correlated systems\cite{sato_current_2010,  wei_enhanced_2019}. Arising from the interplay of strong electron-electron interactions and quasiparticle lifetime renormalization at low temperature, TSTG establishes moiré graphene as a tuneable test bed for heavy fermion physics. Compared to crystalline lattices, TSTG enables a selective tuning between different quantum phases in a single device by a mere change of the electrostatic potentials on the top and bottom gates. Lattice disorder including twist-angle inhomogeneity typically obscures either electronic or thermoelectric properties. Hidden by the percolation to the global probes, some features of interaction-induced quantum states remain elusive to standard magnetotransport measurements. With that, we established a local probe method to investigate strongly-correlated phases by scanning photo-thermoelectric nanoscopy. More experiments are needed to complete the phase diagram shown in the current study. For example, thermopower measurements in the sub-Kelvin range may shed light on the quantum critical behaviour and the emergence of symmetry-broken states, including superconductivity in the moiré lattices\cite{jaoui_quantum_2022}. Our new method applied to the whole class of strongly correlated solid state systems opens a wide range of possibilities to study thermoelectric properties and quantum critical behavior in the clean lattice limit away from the distorted crystalline structures.

\vspace*{1 cm}
\noindent

\vspace*{0.2 cm}
\twocolumngrid

\section*{\textsf{Methods}}
\small
\subsection*{Device fabrication.}
\noindent
The TSTG sample was fabricated using the standard dry-transfer cut-and-stack technique\cite{saito_independent_2020}. A polydimethylsiloxane (PDMS) stamp covered by a poly(bisphenol A carbonate) (PC) film was used to pick up and release the layered single crystals. An AFM tip was used to cut single layer graphene flakes in three parts. By consecutively picking top hBN layer then the top graphene layer, then the graphene middle layer with a target angle of $1.5^\circ$, then a bottom graphene layer twisted back to match the crystallographic axis with the top layer, we achieve a hBN/TSTG heterostructure on the PC layer. At further steps, we pick up the bottom hBN layer to encapsulate TSTG and finally pick up a few layer graphene that serves as a bottom gate to our samples. To prevent a formation of bubbles we preform all transfer steps at 100-110$^\circ$C. At the last step, the heterostructure is released on the target substrate. Choosing the cleanest disorder-free area in the heterostructre, we etch the stack into a Hall-bar shape using a mixed CHF$_3$/O$_2$ plasma. Finally, we couple the sample leads to the metallic electrodes consisted of 10/50 nm layers of Cr/Au.

\subsection*{Cryogenic near-field thermoelectric measurements.}
\noindent
We used a cryogenic scattering-type scanning near-field microscope (cryoSNOM) developed by Neaspec/Attocube to carry our the near-field photovoltage experiments at temperatures between 10-300 K. A tuneable quantum cascade laser (Daylight Solutions) acts as an infrared light source, and the data shown in this work were acquired at an excitation energy of 116 meV (10.6 $\mu$m). We focus approximately 10 mW of light on a PtIr-coated AFM tip (Nanoworld, 50 nm coating),
which is oscillating above the sample surface at $\approx$ 250 kHz with a tapping amplitude of $\approx$ 100 nm. The AFM feedback loop incorporates a system developed by Neaspec/Attocube by which we can lower the quality factor of the AFM cantilever to the values similar to ambient operation at room temperature. A PtIr coated AFM tip is used to scan in tapping mode over the sample in a commercial Neaspec cryoSNOM. The sample voltage is read-out by locking in to the tapping frequency using an Ithaco voltage amplifier and a digital lock-in provided by Neaspec. The resulting photovoltage signal has therefore a sign ambiguity that is resolved by comparing to Mott and theoretical predictions. For the data of this article, the pre-amplifier filter settings were set with a cut-off frequency of 3kHz for the high-pass and around 440kHz for the low-pass (maximum). For this reason, only the 1st harmonic was extracted, which sometimes can lead to far-field contributions that can be removed by a constant offset.

\subsection*{Electronic transport measurement.}
\noindent
Magnetotrasport is measured in a dry commercial cryostat from  Advanced Research Systems (ARS) with ~6 K base temperature and extended dipole magnet of 1 T. Resistivity data is collected in a 4-probe lock-in configuration. We employ a standard low-frequency current-biased lock-in technique with 17.777 Hz excitation frequency and current amplitude of 10 nA. The sample is typically connected in series to the lock-in current output via a 10 M$\Omega$ resistor. The resistance of the sample is 3-4 orders of magnitude smaller than the shunt resistor values providing accurate voltage measurements.

\subsection*{Superlattice charge carrier density extraction.}
We use the relation $n_s = 8 \theta ^2/3a^2$, where $a$ = 0.246 nm is the lattice constant of graphene and $n_s$ is the charge carrier density corresponding to a fully filled superlattice unit cell. We use Hall resistance measurements to identify the back gate capacitance of $C_{BG}$ = 162.54 nF/cm$^2$. Once the gate capacitance is determined, we employ the positions of the longitudinal resistivity peaks (e. g. Fig. ~\ref{Mott breakdown}c) to identify their corresponding charge carrier densities (see e. g. Fig.~\ref{Mott breakdown}c). We further assign the found values to the dimensionless filling factors $\nu$=4$n/n_s$. We highlight here that the same method is used for both Hall density measurements shown in Fig.~\ref{Hall} and for the $PV$ line traces in Fig.~\ref{Comparison}.

\subsection*{Extraction of the photovoltage amplitude at integer fillings.}
To extract the amplitude at integer fillings, we fit a line from filling $\nu$=0.5 to 3.5 and subtract it from the raw data, leveling the background similar to Fig.~\ref{Comparison}b. Then, a digital lock-in is implemented with period $\Delta\nu = 1$ in order to extract the Fourier coefficients of the amplitude. At the final step, the amplitude that computed as the square-root sum of squares. See more detailed information in SI Section V.

\bibliographystyle{naturemag-sergi}
\bibliography{References.bib}

\begin{thebibliography}{10}
\expandafter\ifx\csname url\endcsname\relax
  \def\url#1{\texttt{#1}}\fi
\expandafter\ifx\csname urlprefix\endcsname\relax\def\urlprefix{}\fi
\providecommand{\bibinfo}[2]{#2}
\providecommand{\eprint}[2][]{\url{#2}}

\bibitem{wirth_exploring_2016}
\bibinfo{author}{Wirth, S.} \& \bibinfo{author}{Steglich, F.}
\newblock \href{http://dx.doi.org/10.1038/natrevmats.2016.51}{Exploring heavy
  fermions from macroscopic to microscopic length scales}.
\newblock \textit{\bibinfo{journal}{Nature Reviews Materials}}
  \textbf{\bibinfo{volume}{1}}, \bibinfo{pages}{1--16} (\bibinfo{year}{2016}).

\bibitem{stewart_heavy-fermion_1984}
\bibinfo{author}{Stewart, S.~G.}
\newblock \href{http://dx.doi.org/10.1103/RevModPhys.56.755}{Heavy-fermion
  systems}.
\newblock \textit{\bibinfo{journal}{Reviews of Modern Physics}}
  \textbf{\bibinfo{volume}{56}}, \bibinfo{pages}{755} (\bibinfo{year}{1984}).

\bibitem{landau_lev_davidovich_theory_1956}
\bibinfo{author}{Landau, L.~D.}
\newblock Theory of {Fermi}-{Liquids}.
\newblock \textit{\bibinfo{journal}{Soviet Physics JETP-USSR}}
  \textbf{\bibinfo{volume}{3}}, \bibinfo{pages}{920--925}
  (\bibinfo{year}{1956}).

\bibitem{andres_4_1975}
\bibinfo{author}{Andres, K.}, \bibinfo{author}{Graebner, J.} \&
  \bibinfo{author}{Ott, H.}
\newblock \href{http://dx.doi.org/10.1103/PhysRevLett.35.1779}{4
  f-{Virtual}-{Bound}-{State} {Formation} in {Ce} {Al} 3 at {Low}
  {Temperatures}}.
\newblock \textit{\bibinfo{journal}{Physical Review Letters}}
  \textbf{\bibinfo{volume}{35}}, \bibinfo{pages}{1779} (\bibinfo{year}{1975}).

\bibitem{steglich_superconductivity_1979}
\bibinfo{author}{Steglich, F.} et~al.
\newblock
  \href{http://dx.doi.org/10.1103/PhysRevLett.43.1892}{Superconductivity in the
  presence of strong pauli paramagnetism: {Ce} cu 2 si 2}.
\newblock \textit{\bibinfo{journal}{Physical Review Letters}}
  \textbf{\bibinfo{volume}{43}}, \bibinfo{pages}{1892} (\bibinfo{year}{1979}).

\bibitem{radovan_magnetic_2003}
\bibinfo{author}{Radovan, H.} et~al.
\newblock \href{http://dx.doi.org/10.1038/nature01842}{Magnetic enhancement of
  superconductivity from electron spin domains}.
\newblock \textit{\bibinfo{journal}{Nature}} \textbf{\bibinfo{volume}{425}},
  \bibinfo{pages}{51--55} (\bibinfo{year}{2003}).

\bibitem{jonson_mott_1980}
\bibinfo{author}{Jonson, M.} \& \bibinfo{author}{Mahan, G.~D.}
\newblock \href{http://dx.doi.org/10.1103/PhysRevB.21.4223}{Mott's formula for
  the thermopower and the wiedemann-franz law}.
\newblock \textit{\bibinfo{journal}{Phys. Rev. B}}
  \textbf{\bibinfo{volume}{21}}, \bibinfo{pages}{4223--4229}
  (\bibinfo{year}{1980}).

\bibitem{bistritzer_moire_2011}
\bibinfo{author}{Bistritzer, R.} \& \bibinfo{author}{MacDonald, A.~H.}
\newblock \href{http://dx.doi.org/10.1073/pnas.1108174108}{Moiré bands in
  twisted double-layer graphene}.
\newblock \textit{\bibinfo{journal}{Proceedings of the National Academy of
  Sciences}} \textbf{\bibinfo{volume}{108}}, \bibinfo{pages}{12233--12237}
  (\bibinfo{year}{2011}).

\bibitem{andrei_graphene_2020}
\bibinfo{author}{Andrei, E.~Y.} \& \bibinfo{author}{MacDonald, A.~H.}
\newblock \href{http://dx.doi.org/10.1038/s41563-020-00840-0}{Graphene bilayers
  with a twist}.
\newblock \textit{\bibinfo{journal}{Nature Materials}}
  \textbf{\bibinfo{volume}{19}}, \bibinfo{pages}{1265--1275}
  (\bibinfo{year}{2020}).

\bibitem{balents_superconductivity_2020}
\bibinfo{author}{Balents, L.}, \bibinfo{author}{Dean, C.~R.},
  \bibinfo{author}{Efetov, D.~K.} \& \bibinfo{author}{Young, A.~F.}
\newblock \href{http://dx.doi.org/10.1038/s41567-020-0906-9}{Superconductivity
  and strong correlations in moiré flat bands}.
\newblock \textit{\bibinfo{journal}{Nature Physics}}
  \textbf{\bibinfo{volume}{16}}, \bibinfo{pages}{725--733}
  (\bibinfo{year}{2020}).

\bibitem{cao_unconventional_2018}
\bibinfo{author}{Cao, Y.} et~al.
\newblock \href{http://dx.doi.org/10.1038/nature26160}{Unconventional
  superconductivity in magic-angle graphene superlattices}.
\newblock \textit{\bibinfo{journal}{Nature}} \textbf{\bibinfo{volume}{556}},
  \bibinfo{pages}{43--50} (\bibinfo{year}{2018}).

\bibitem{yankowitz_tuning_2019}
\bibinfo{author}{Yankowitz, M.} et~al.
\newblock \href{http://dx.doi.org/10.1126/science.aav1910}{Tuning
  superconductivity in twisted bilayer graphene}.
\newblock \textit{\bibinfo{journal}{Science}} \textbf{\bibinfo{volume}{363}},
  \bibinfo{pages}{1059--1064} (\bibinfo{year}{2019}).

\bibitem{lu_superconductors_2019}
\bibinfo{author}{Lu, X.} et~al.
\newblock \href{http://dx.doi.org/10.1038/s41586-019-1695-0}{Superconductors,
  orbital magnets and correlated states in magic-angle bilayer graphene}.
\newblock \textit{\bibinfo{journal}{Nature}} \textbf{\bibinfo{volume}{574}},
  \bibinfo{pages}{653--657} (\bibinfo{year}{2019}).

\bibitem{po_origin_2018}
\bibinfo{author}{Po, H.~C.}, \bibinfo{author}{Zou, L.},
  \bibinfo{author}{Vishwanath, A.} \& \bibinfo{author}{Senthil, T.}
\newblock \href{http://dx.doi.org/10.1103/PhysRevX.8.031089}{Origin of {Mott}
  insulating behavior and superconductivity in twisted bilayer graphene}.
\newblock \textit{\bibinfo{journal}{Physical Review X}}
  \textbf{\bibinfo{volume}{8}}, \bibinfo{pages}{031089} (\bibinfo{year}{2018}).

\bibitem{dodaro_phases_2018}
\bibinfo{author}{Dodaro, J.~F.}, \bibinfo{author}{Kivelson, S.~A.},
  \bibinfo{author}{Schattner, Y.}, \bibinfo{author}{Sun, X.-Q.} \&
  \bibinfo{author}{Wang, C.}
\newblock \href{http://dx.doi.org/10.1103/PhysRevB.98.075154}{Phases of a
  phenomenological model of twisted bilayer graphene}.
\newblock \textit{\bibinfo{journal}{Physical Review B}}
  \textbf{\bibinfo{volume}{98}}, \bibinfo{pages}{075154}
  (\bibinfo{year}{2018}).

\bibitem{xie_nature_2020}
\bibinfo{author}{Xie, M.} \& \bibinfo{author}{MacDonald, A.~H.}
\newblock \href{http://dx.doi.org/10.1103/PhysRevLett.124.097601}{Nature of the
  {Correlated} {Insulator} {States} in {Twisted} {Bilayer} {Graphene}}.
\newblock \textit{\bibinfo{journal}{Physical Review Letters}}
  \textbf{\bibinfo{volume}{124}}, \bibinfo{pages}{097601}
  (\bibinfo{year}{2020}).

\bibitem{oh_evidence_2021}
\bibinfo{author}{Oh, M.} et~al.
\newblock \href{http://dx.doi.org/10.1038/s41586-021-04121-x}{Evidence for
  unconventional superconductivity in twisted bilayer graphene}.
\newblock \textit{\bibinfo{journal}{Nature}} \textbf{\bibinfo{volume}{600}},
  \bibinfo{pages}{240--245} (\bibinfo{year}{2021}).

\bibitem{cao_correlated_2018}
\bibinfo{author}{Cao, Y.} et~al.
\newblock \href{http://dx.doi.org/10.1038/nature26154}{Correlated insulator
  behaviour at half-filling in magic-angle graphene superlattices}.
\newblock \textit{\bibinfo{journal}{Nature}} \textbf{\bibinfo{volume}{556}},
  \bibinfo{pages}{80--84} (\bibinfo{year}{2018}).

\bibitem{nuckolls_strongly_2020}
\bibinfo{author}{Nuckolls, K.~P.} et~al.
\newblock \href{http://dx.doi.org/10.1038/s41586-020-3028-8}{Strongly
  correlated {Chern} insulators in magic-angle twisted bilayer graphene}.
\newblock \textit{\bibinfo{journal}{Nature}} \textbf{\bibinfo{volume}{588}},
  \bibinfo{pages}{610--615} (\bibinfo{year}{2020}).

\bibitem{song_topology_2019}
\bibinfo{author}{Song, Z.} et~al.
\newblock \href{http://dx.doi.org/10.1103/PhysRevLett.123.036401}{All {{Magic
  Angles}} in {{Twisted Bilayer Graphene}} are {{Topological}}}.
\newblock \textit{\bibinfo{journal}{Phys. Rev. Lett.}}
  \textbf{\bibinfo{volume}{123}}, \bibinfo{pages}{036401}
  (\bibinfo{year}{2019}).

\bibitem{wagner_global_2022}
\bibinfo{author}{Wagner, G.}, \bibinfo{author}{Kwan, Y.~H.},
  \bibinfo{author}{Bultinck, N.}, \bibinfo{author}{Simon, S.~H.} \&
  \bibinfo{author}{Parameswaran, S.}
\newblock \href{http://dx.doi.org/10.1103/PhysRevLett.128.156401}{Global phase
  diagram of the normal state of twisted bilayer graphene}.
\newblock \textit{\bibinfo{journal}{Physical Review Letters}}
  \textbf{\bibinfo{volume}{128}}, \bibinfo{pages}{156401}
  (\bibinfo{year}{2022}).

\bibitem{lian_twisted_2021}
\bibinfo{author}{Lian, B.} et~al.
\newblock \href{http://dx.doi.org/10.1103/PhysRevB.103.205414}{Twisted bilayer
  graphene. {IV}. {Exact} insulator ground states and phase diagram}.
\newblock \textit{\bibinfo{journal}{Physical Review B}}
  \textbf{\bibinfo{volume}{103}}, \bibinfo{pages}{205414}
  (\bibinfo{year}{2021}).

\bibitem{wu_chern_2021}
\bibinfo{author}{Wu, S.}, \bibinfo{author}{Zhang, Z.},
  \bibinfo{author}{Watanabe, K.}, \bibinfo{author}{Taniguchi, T.} \&
  \bibinfo{author}{Andrei, E.~Y.}
\newblock \href{http://dx.doi.org/10.1038/s41563-020-00911-2}{Chern insulators,
  van {Hove} singularities and topological flat bands in magic-angle twisted
  bilayer graphene}.
\newblock \textit{\bibinfo{journal}{Nature materials}}
  \textbf{\bibinfo{volume}{20}}, \bibinfo{pages}{488--494}
  (\bibinfo{year}{2021}).

\bibitem{serlin_intrinsic_2020}
\bibinfo{author}{Serlin, M.} et~al.
\newblock \href{http://dx.doi.org/10.1126/science.aay5533}{Intrinsic quantized
  anomalous {Hall} effect in a moiré heterostructure}.
\newblock \textit{\bibinfo{journal}{Science}} \textbf{\bibinfo{volume}{367}},
  \bibinfo{pages}{900--903} (\bibinfo{year}{2020}).

\bibitem{sharpe_emergent_2019}
\bibinfo{author}{Sharpe, A.~L.} et~al.
\newblock \href{http://dx.doi.org/10.1126/science.aaw3780}{Emergent
  ferromagnetism near three-quarters filling in twisted bilayer graphene}.
\newblock \textit{\bibinfo{journal}{Science}} \textbf{\bibinfo{volume}{365}},
  \bibinfo{pages}{605--608} (\bibinfo{year}{2019}).

\bibitem{polshyn_electrical_2020}
\bibinfo{author}{Polshyn, H.} et~al.
\newblock \href{http://dx.doi.org/10.1038/s41586-020-2963-8}{Electrical
  switching of magnetic order in an orbital {Chern} insulator}.
\newblock \textit{\bibinfo{journal}{Nature}} \textbf{\bibinfo{volume}{588}},
  \bibinfo{pages}{66--70} (\bibinfo{year}{2020}).

\bibitem{tang_simulation_2020}
\bibinfo{author}{Tang, Y.} et~al.
\newblock \href{http://dx.doi.org/10.1038/s41586-020-2085-3}{Simulation of
  {Hubbard} model physics in {WSe2}/{WS2} moiré superlattices}.
\newblock \textit{\bibinfo{journal}{Nature}} \textbf{\bibinfo{volume}{579}},
  \bibinfo{pages}{353--358} (\bibinfo{year}{2020}).

\bibitem{zhao_gate-tunable_2023}
\bibinfo{author}{Zhao, W.} et~al.
\newblock \href{http://dx.doi.org/10.1038/s41586-023-05800-7}{Gate-tunable
  heavy fermions in a moiré {Kondo} lattice}.
\newblock \textit{\bibinfo{journal}{Nature}} \textbf{\bibinfo{volume}{616}},
  \bibinfo{pages}{61--65} (\bibinfo{year}{2023}).

\bibitem{park_tunable_2021}
\bibinfo{author}{Park, J.~M.}, \bibinfo{author}{Cao, Y.},
  \bibinfo{author}{Watanabe, K.}, \bibinfo{author}{Taniguchi, T.} \&
  \bibinfo{author}{Jarillo-Herrero, P.}
\newblock \href{http://dx.doi.org/10.1038/s41586-021-03192-0}{Tunable strongly
  coupled superconductivity in magic-angle twisted trilayer graphene}.
\newblock \textit{\bibinfo{journal}{Nature}} \textbf{\bibinfo{volume}{590}},
  \bibinfo{pages}{249--255} (\bibinfo{year}{2021}).

\bibitem{cao_pauli-limit_2021}
\bibinfo{author}{Cao, Y.}, \bibinfo{author}{Park, J.~M.},
  \bibinfo{author}{Watanabe, K.}, \bibinfo{author}{Taniguchi, T.} \&
  \bibinfo{author}{Jarillo-Herrero, P.}
\newblock \href{http://dx.doi.org/10.1038/s41586-021-03685-y}{Pauli-limit
  violation and re-entrant superconductivity in moiré graphene}.
\newblock \textit{\bibinfo{journal}{Nature}} \textbf{\bibinfo{volume}{595}},
  \bibinfo{pages}{526--531} (\bibinfo{year}{2021}).

\bibitem{jiang_charge_2019}
\bibinfo{author}{Jiang, Y.} et~al.
\newblock \href{http://dx.doi.org/10.1038/s41586-019-1460-4}{Charge order and
  broken rotational symmetry in magic-angle twisted bilayer graphene}.
\newblock \textit{\bibinfo{journal}{Nature}} \textbf{\bibinfo{volume}{573}},
  \bibinfo{pages}{91--95} (\bibinfo{year}{2019}).

\bibitem{kerelsky_maximized_2019}
\bibinfo{author}{Kerelsky, A.} et~al.
\newblock \href{http://dx.doi.org/10.1038/s41586-019-1431-9}{Maximized electron
  interactions at the magic angle in twisted bilayer graphene}.
\newblock \textit{\bibinfo{journal}{Nature}} \textbf{\bibinfo{volume}{572}},
  \bibinfo{pages}{95--100} (\bibinfo{year}{2019}).

\bibitem{wong_cascade_2020}
\bibinfo{author}{Wong, D.} et~al.
\newblock \href{http://dx.doi.org/10.1038/s41586-020-2339-0}{Cascade of
  electronic transitions in magic-angle twisted bilayer graphene}.
\newblock \textit{\bibinfo{journal}{Nature}} \textbf{\bibinfo{volume}{582}},
  \bibinfo{pages}{198--202} (\bibinfo{year}{2020}).

\bibitem{saito_isospin_2021}
\bibinfo{author}{Saito, Y.} et~al.
\newblock \href{http://dx.doi.org/10.1038/s41586-021-03409-2}{Isospin
  {Pomeranchuk} effect in twisted bilayer graphene}.
\newblock \textit{\bibinfo{journal}{Nature}} \textbf{\bibinfo{volume}{592}},
  \bibinfo{pages}{220--224} (\bibinfo{year}{2021}).

\bibitem{zondiner_cascade_2020}
\bibinfo{author}{Zondiner, U.} et~al.
\newblock \href{http://dx.doi.org/10.1038/s41586-020-2373-y}{Cascade of phase
  transitions and {Dirac} revivals in magic-angle graphene}.
\newblock \textit{\bibinfo{journal}{Nature}} \textbf{\bibinfo{volume}{582}},
  \bibinfo{pages}{203--208} (\bibinfo{year}{2020}).

\bibitem{rozen_entropic_2021}
\bibinfo{author}{Rozen, A.} et~al.
\newblock \href{http://dx.doi.org/10.1038/s41586-021-03319-3}{Entropic evidence
  for a {Pomeranchuk} effect in magic-angle graphene}.
\newblock \textit{\bibinfo{journal}{Nature}} \textbf{\bibinfo{volume}{592}},
  \bibinfo{pages}{214--219} (\bibinfo{year}{2021}).

\bibitem{choi_correlation-driven_2021}
\bibinfo{author}{Choi, Y.} et~al.
\newblock
  \href{http://dx.doi.org/10.1038/s41586-020-03159-7}{Correlation-driven
  topological phases in magic-angle twisted bilayer graphene}.
\newblock \textit{\bibinfo{journal}{Nature}} \textbf{\bibinfo{volume}{589}},
  \bibinfo{pages}{536--541} (\bibinfo{year}{2021}).

\bibitem{calugaru_seebeck}
\bibinfo{author}{C\u{a}lug\u{a}ru, D.} et~al.
\newblock The thermoelectric effect and its natural heavy fermion explanation
  in twisted bilayer and trilayer graphene.
\newblock \textit{\bibinfo{journal}{preprint to appear}}
  (\bibinfo{year}{2024}).

\bibitem{song_magic-angle_2022}
\bibinfo{author}{Song, Z.-D.} \& \bibinfo{author}{Bernevig, B.~A.}
\newblock \href{http://dx.doi.org/10.1103/PhysRevLett.129.047601}{Magic-{Angle}
  {Twisted} {Bilayer} {Graphene} as a {Topological} {Heavy} {Fermion}
  {Problem}}.
\newblock \textit{\bibinfo{journal}{Physical Review Letters}}
  \textbf{\bibinfo{volume}{129}}, \bibinfo{pages}{047601}
  (\bibinfo{year}{2022}).

\bibitem{yu_magic-angle_2023}
\bibinfo{author}{Yu, J.}, \bibinfo{author}{Xie, M.}, \bibinfo{author}{Bernevig,
  B.~A.} \& \bibinfo{author}{Sarma, S.~D.}
\newblock \href{http://dx.doi.org/10.1103/PhysRevB.108.035129}{Magic-{Angle}
  {Twisted} {Symmetric} {Trilayer} {Graphene} as {Topological} {Heavy}
  {Fermion} {Problem}}.
\newblock \textit{\bibinfo{journal}{Physical Review B}}
  \textbf{\bibinfo{volume}{108}}, \bibinfo{pages}{035129}
  (\bibinfo{year}{2023}).

\bibitem{woessner_near-field_2016}
\bibinfo{author}{Woessner, A.} et~al.
\newblock \href{http://dx.doi.org/10.1038/ncomms10783}{Near-field photocurrent
  nanoscopy on bare and encapsulated graphene}.
\newblock \textit{\bibinfo{journal}{Nature Communications}}
  \textbf{\bibinfo{volume}{7}}, \bibinfo{pages}{10783} (\bibinfo{year}{2016}).

\bibitem{kim_evidence_2023}
\bibinfo{author}{Kim, H.} et~al.
\newblock
  \href{http://dx.doi.org/https://doi.org/10.1038/s41586-022-04715-z}{Evidence
  for unconventional superconductivity in twisted trilayer graphene}.
\newblock \textit{\bibinfo{journal}{Nature}} \textbf{\bibinfo{volume}{606}},
  \bibinfo{pages}{494--500} (\bibinfo{year}{2022}).

\bibitem{ghawri2022breakdown}
\bibinfo{author}{Ghawri, B.} et~al.
\newblock
  \href{http://dx.doi.org/https://doi.org/10.1038/s41467-022-29198-4}{Breakdown
  of semiclassical description of thermoelectricity in near-magic angle twisted
  bilayer graphene}.
\newblock \textit{\bibinfo{journal}{Nature Communications}}
  \textbf{\bibinfo{volume}{13}}, \bibinfo{pages}{1522} (\bibinfo{year}{2022}).

\bibitem{gabor_hot_2011}
\bibinfo{author}{Gabor, N.~M.} et~al.
\newblock \href{http://dx.doi.org/10.1126/science.1211384}{Hot
  {Carrier}-{Assisted} {Intrinsic} {Photoresponse} in {Graphene}}.
\newblock \textit{\bibinfo{journal}{Science}} \textbf{\bibinfo{volume}{334}},
  \bibinfo{pages}{648--652} (\bibinfo{year}{2011}).

\bibitem{ma_tuning_2016}
\bibinfo{author}{Ma, Q.} et~al.
\newblock \href{http://dx.doi.org/10.1038/nphys3620}{Tuning ultrafast electron
  thermalization pathways in a van der {Waals} heterostructure}.
\newblock \textit{\bibinfo{journal}{Nature Physics}}
  \textbf{\bibinfo{volume}{12}}, \bibinfo{pages}{455--459}
  (\bibinfo{year}{2016}).

\bibitem{dresselhaus2007thermoelectricity}
\bibinfo{author}{Dresselhaus, M.} et~al.
\newblock \href{http://dx.doi.org/https://doi.org/10.1002/adma.200600527}{New
  directions for low-dimensional thermoelectric materials}.
\newblock \textit{\bibinfo{journal}{Advanced Materials}}
  \textbf{\bibinfo{volume}{19}}, \bibinfo{pages}{1043--1053}
  (\bibinfo{year}{2007}).

\bibitem{paul_interaction-driven_2022}
\bibinfo{author}{Paul, A.~K.} et~al.
\newblock
  \href{http://dx.doi.org/10.1038/s41567-022-01574-3}{Interaction-driven giant
  thermopower in magic-angle twisted bilayer graphene}.
\newblock \textit{\bibinfo{journal}{Nature Physics}}
  \textbf{\bibinfo{volume}{18}}, \bibinfo{pages}{691--698}
  (\bibinfo{year}{2022}).

\bibitem{yu_TSTG_THF_2023}
\bibinfo{author}{Yu, J.}, \bibinfo{author}{Xie, M.}, \bibinfo{author}{Bernevig,
  B.~A.} \& \bibinfo{author}{Sarma, S.~D.}
\newblock Magic-angle twisted symmetric trilayer graphene as topological heavy
  fermion problem.
\newblock \textit{\bibinfo{journal}{arXiv preprint arXiv:2301.04171}}
  (\bibinfo{year}{2023}).

\bibitem{datta_heavy_2023}
\bibinfo{author}{Datta, A.}, \bibinfo{author}{Calderón, M.~J.},
  \bibinfo{author}{Camjayi, A.} \& \bibinfo{author}{Bascones, E.}
\newblock \href{http://dx.doi.org/10.1038/s41467-023-40754-4}{Heavy
  quasiparticles and cascades without symmetry breaking in twisted bilayer
  graphene}.
\newblock \textit{\bibinfo{journal}{Nature Communications}}
  \textbf{\bibinfo{volume}{14}}, \bibinfo{pages}{5036} (\bibinfo{year}{2023}).

\bibitem{novelli_optical_2020}
\bibinfo{author}{Novelli, P.}, \bibinfo{author}{Torre, I.},
  \bibinfo{author}{Koppens, F. H.~L.}, \bibinfo{author}{Taddei, F.} \&
  \bibinfo{author}{Polini, M.}
\newblock \href{http://dx.doi.org/10.1103/PhysRevB.102.125403}{Optical and
  plasmonic properties of twisted bilayer graphene: {Impact} of interlayer
  tunneling asymmetry and ground-state charge inhomogeneity}.
\newblock \textit{\bibinfo{journal}{Physical Review B}}
  \textbf{\bibinfo{volume}{102}}, \bibinfo{pages}{125403}
  (\bibinfo{year}{2020}).

\bibitem{hu_kondo_2023}
\bibinfo{author}{Hu, H.}, \bibinfo{author}{Bernevig, B.~A.} \&
  \bibinfo{author}{Tsvelik, A.~M.}
\newblock \href{http://dx.doi.org/10.1103/PhysRevLett.131.026502}{Kondo lattice
  model of magic-angle twisted-bilayer graphene: {Hund}’s rule, local-moment
  fluctuations, and low-energy effective theory}.
\newblock \textit{\bibinfo{journal}{Physical Review Letters}}
  \textbf{\bibinfo{volume}{131}}, \bibinfo{pages}{026502}
  (\bibinfo{year}{2023}).

\bibitem{lin_zero-field_2022}
\bibinfo{author}{Lin, J.-X.} et~al.
\newblock \href{http://dx.doi.org/10.1038/s41567-022-01700-1}{Zero-field
  superconducting diode effect in small-twist-angle trilayer graphene}.
\newblock \textit{\bibinfo{journal}{Nature Physics}}
  \textbf{\bibinfo{volume}{18}}, \bibinfo{pages}{1221--1227}
  (\bibinfo{year}{2022}).

\bibitem{Bennett_2023}
\bibinfo{author}{Bennett, D.}, \bibinfo{author}{Larson, D.~T.},
  \bibinfo{author}{Sharma, L.}, \bibinfo{author}{Carr, S.} \&
  \bibinfo{author}{Kaxiras, E.}
\newblock Twisted bilayer graphene revisited: minimal two-band model for
  low-energy bands.
\newblock \textit{\bibinfo{journal}{arXiv preprint arXiv:2310.12308}}
  (\bibinfo{year}{2023}).

\bibitem{sato_current_2010}
\bibinfo{author}{Sato, H.} et~al.
\newblock \href{http://dx.doi.org/10.1016/j.physc.2009.11.176}{Current
  understanding of the heavy fermion superconductivity in {PrOs4Sb12}}.
\newblock \textit{\bibinfo{journal}{Physica C: Superconductivity and its
  Applications}} \textbf{\bibinfo{volume}{470}}, \bibinfo{pages}{S525--S528}
  (\bibinfo{year}{2010}).

\bibitem{wei_enhanced_2019}
\bibinfo{author}{Wei, K.} et~al.
\newblock \href{http://dx.doi.org/10.1126/sciadv.aaw6183}{Enhanced
  thermoelectric performance of heavy-fermion compounds {YbTM2Zn20} ({TM} =
  {Co}, {Rh}, {Ir}) at low temperatures}.
\newblock \textit{\bibinfo{journal}{Science Advances}}
  \textbf{\bibinfo{volume}{5}}, \bibinfo{pages}{eaaw6183}
  (\bibinfo{year}{2019}).

\bibitem{jaoui_quantum_2022}
\bibinfo{author}{Jaoui, A.} et~al.
\newblock \href{http://dx.doi.org/10.1038/s41567-022-01556-5}{Quantum critical
  behaviour in magic-angle twisted bilayer graphene}.
\newblock \textit{\bibinfo{journal}{Nature Physics}}
  \textbf{\bibinfo{volume}{18}}, \bibinfo{pages}{633--638}
  (\bibinfo{year}{2022}).

\bibitem{saito_independent_2020}
\bibinfo{author}{Saito, Y.}, \bibinfo{author}{Ge, J.},
  \bibinfo{author}{Watanabe, K.}, \bibinfo{author}{Taniguchi, T.} \&
  \bibinfo{author}{Young, A.~F.}
\newblock \href{http://dx.doi.org/10.1038/s41567-020-0928-3}{Independent
  superconductors and correlated insulators in twisted bilayer graphene}.
\newblock \textit{\bibinfo{journal}{Nature Physics}}
  \textbf{\bibinfo{volume}{16}}, \bibinfo{pages}{926--930}
  (\bibinfo{year}{2020}).

\end{thebibliography}


\begin{thebibliography}{10}
\expandafter\ifx\csname url\endcsname\relax
  \def\url#1{\texttt{#1}}\fi
\expandafter\ifx\csname urlprefix\endcsname\relax\def\urlprefix{}\fi
\providecommand{\bibinfo}[2]{#2}
\providecommand{\eprint}[2][]{\url{#2}}

\bibitem{woessner_near-field_2016}
\bibinfo{author}{Woessner, A.} et~al.
\newblock \href{http://dx.doi.org/10.1038/ncomms10783}{Near-field photocurrent
  nanoscopy on bare and encapsulated graphene}.
\newblock \textit{\bibinfo{journal}{Nature Communications}}
  \textbf{\bibinfo{volume}{7}}, \bibinfo{pages}{10783} (\bibinfo{year}{2016}).

\bibitem{gabor_hot_2011}
\bibinfo{author}{Gabor, N.~M.} et~al.
\newblock \href{http://dx.doi.org/10.1126/science.1211384}{Hot
  {Carrier}-{Assisted} {Intrinsic} {Photoresponse} in {Graphene}}.
\newblock \textit{\bibinfo{journal}{Science}} \textbf{\bibinfo{volume}{334}},
  \bibinfo{pages}{648--652} (\bibinfo{year}{2011}).

\bibitem{jonson_mott_1980}
\bibinfo{author}{Jonson, M.} \& \bibinfo{author}{Mahan, G.~D.}
\newblock \href{http://dx.doi.org/10.1103/PhysRevB.21.4223}{Mott's formula for
  the thermopower and the wiedemann-franz law}.
\newblock \textit{\bibinfo{journal}{Phys. Rev. B}}
  \textbf{\bibinfo{volume}{21}}, \bibinfo{pages}{4223--4229}
  (\bibinfo{year}{1980}).

\bibitem{calugaru_seebeck}
\bibinfo{author}{C\u{a}lug\u{a}ru, D.} et~al.
\newblock The thermoelectric effect and its natural heavy fermion explanation
  in twisted bilayer and trilayer graphene.
\newblock \textit{\bibinfo{journal}{preprint to appear}}
  (\bibinfo{year}{2024}).

\bibitem{novelli_optical_2020}
\bibinfo{author}{Novelli, P.}, \bibinfo{author}{Torre, I.},
  \bibinfo{author}{Koppens, F. H.~L.}, \bibinfo{author}{Taddei, F.} \&
  \bibinfo{author}{Polini, M.}
\newblock \href{http://dx.doi.org/10.1103/PhysRevB.102.125403}{Optical and
  plasmonic properties of twisted bilayer graphene: {Impact} of interlayer
  tunneling asymmetry and ground-state charge inhomogeneity}.
\newblock \textit{\bibinfo{journal}{Physical Review B}}
  \textbf{\bibinfo{volume}{102}}, \bibinfo{pages}{125403}
  (\bibinfo{year}{2020}).

\bibitem{hu_kondo_2023}
\bibinfo{author}{Hu, H.}, \bibinfo{author}{Bernevig, B.~A.} \&
  \bibinfo{author}{Tsvelik, A.~M.}
\newblock \href{http://dx.doi.org/10.1103/PhysRevLett.131.026502}{Kondo lattice
  model of magic-angle twisted-bilayer graphene: {Hund}’s rule, local-moment
  fluctuations, and low-energy effective theory}.
\newblock \textit{\bibinfo{journal}{Physical Review Letters}}
  \textbf{\bibinfo{volume}{131}}, \bibinfo{pages}{026502}
  (\bibinfo{year}{2023}).

\bibitem{song_magic-angle_2022}
\bibinfo{author}{Song, Z.-D.} \& \bibinfo{author}{Bernevig, B.~A.}
\newblock \href{http://dx.doi.org/10.1103/PhysRevLett.129.047601}{Magic-{Angle}
  {Twisted} {Bilayer} {Graphene} as a {Topological} {Heavy} {Fermion}
  {Problem}}.
\newblock \textit{\bibinfo{journal}{Physical Review Letters}}
  \textbf{\bibinfo{volume}{129}}, \bibinfo{pages}{047601}
  (\bibinfo{year}{2022}).

\bibitem{calugaru_tbg_2023}
\bibinfo{author}{Călugăru, D.} et~al.
\newblock \href{http://dx.doi.org/10.1063/10.0019421}{{TBG} as {Topological}
  {Heavy} {Fermion}: {II}. {Analytical} approximations of the model
  parameters}.
\newblock \textit{\bibinfo{journal}{Low Temperature Physics}}
  \textbf{\bibinfo{volume}{49}}, \bibinfo{pages}{640--654}
  (\bibinfo{year}{2023}).

\bibitem{cao_correlated_2018}
\bibinfo{author}{Cao, Y.} et~al.
\newblock \href{http://dx.doi.org/10.1038/nature26154}{Correlated insulator
  behaviour at half-filling in magic-angle graphene superlattices}.
\newblock \textit{\bibinfo{journal}{Nature}} \textbf{\bibinfo{volume}{556}},
  \bibinfo{pages}{80--84} (\bibinfo{year}{2018}).

\bibitem{song_shockley-ramo_2014}
\bibinfo{author}{Song, J. C.~W.} \& \bibinfo{author}{Levitov, L.~S.}
\newblock \href{http://dx.doi.org/10.1103/PhysRevB.90.075415}{Shockley-{Ramo}
  theorem and long-range photocurrent response in gapless materials}.
\newblock \textit{\bibinfo{journal}{Physical Review B}}
  \textbf{\bibinfo{volume}{90}}, \bibinfo{pages}{075415}
  (\bibinfo{year}{2014}).

\bibitem{zhao_gate-tunable_2023}
\bibinfo{author}{Zhao, W.} et~al.
\newblock \href{http://dx.doi.org/10.1038/s41586-023-05800-7}{Gate-tunable
  heavy fermions in a moiré {Kondo} lattice}.
\newblock \textit{\bibinfo{journal}{Nature}} \textbf{\bibinfo{volume}{616}},
  \bibinfo{pages}{61--65} (\bibinfo{year}{2023}).

\bibitem{ramires_emulating_2021}
\bibinfo{author}{Ramires, A.} \& \bibinfo{author}{Lado, J.~L.}
\newblock \href{http://dx.doi.org/10.1103/PhysRevLett.127.026401}{Emulating
  {Heavy} {Fermions} in {Twisted} {Trilayer} {Graphene}}.
\newblock \textit{\bibinfo{journal}{Physical Review Letters}}
  \textbf{\bibinfo{volume}{127}}, \bibinfo{pages}{026401}
  (\bibinfo{year}{2021}).

\bibitem{hesp_cryogenic_2023}
\bibinfo{author}{Hesp, N. C.~H.} et~al.
\newblock Cryogenic nano-imaging of second-order moire superlattices.
\newblock \textit{\bibinfo{journal}{arXiv preprint arXiv:2302.05487}}
  (\bibinfo{year}{2023}).

\end{thebibliography}

\vspace*{-0.2cm}
\section*{\textsf{Acknowledgements}}
\vspace*{-0.2cm}
\noindent

We thank for helpful discussions with Justin C. W. Song, Allan H. MacDonald, Pablo Jarillo-Herrero, Stephan Roche and Dmitri K. Efetov. F.H.L.K. acknowledges support from the ERC TOPONANOP (726001),  the government of Spain (PID2019-106875GB-I00; Severo Ochoa CEX2019-000910-S [MCIN/AEI/10.13039/501100011033], PCI2021-122020-2A funded by MCIN/AEI/ 10.13039/501100011033), the "European Union NextGenerationEU/PRTR (PRTR-C17.I1), Fundació Cellex, Fundació Mir-Puig, and Generalitat de Catalunya (CERCA, AGAUR, 2021 SGR 01443). Furthermore, the research leading to these results has received funding from the European Union’s Horizon 2020  under grant agreement  no. 881603 (Graphene flagship Core3) and 820378 (Quantum flagship). This material is based upon work supported by the Air Force Office of Scientific Research under award number FA8655-23-1-7047. Any opinions, findings, and conclusions or recommendations expressed in this material are those of the author(s) and do not necessarily reflect the views of the United States Air Force. 
D.C. acknowledges the hospitality of the Donostia International Physics Center, at which this work was carried out. B.A.B. was supported by DOE Grant No. DE-SC0016239. D.C. was supported by the European Research Council (ERC) under the European Union’s Horizon 2020 research and innovation program (grant agreement no. 101020833) and by the Simons Investigator Grant No. 404513, the Gordon and Betty Moore Foundation through Grant No. GBMF8685 towards the Princeton theory program, the Gordon and Betty Moore Foundation’s EPiQS Initiative (Grant No. GBMF11070), Office of Naval Research (ONR Grant No. N00014-20-1-2303), Global Collaborative Network Grant at Princeton University, BSF Israel US foundation No. 2018226, NSF-MERSEC (Grant No. MERSEC DMR 2011750). H.H. was supported by the European Research Council (ERC) under the European Union’s Horizon 2020 research and innovation program (Grant Agreement No. 101020833) and the Schmidt Fund Grant.
P.S. acknowledges support from the European Union’s Horizon 2020 research and innovation programme under the Marie Skłodowska-Curie Grant No. 754510.
S.B.P. acknowledges funding from the “Presidencia de la Agencia Estatal de Investigación” within the PRE2020-094404 predoctoral fellowship.
N.C.H.H. acknowledges funding from the European Union's Horizon 2020 research and innovation programme under the Marie Skłodowska-Curie grant agreement Ref. 665884.
K.W. and T.T. acknowledge support from JSPS KAKENHI (Grant Refs. 19H05790, 20H00354 and 21H05233).

\vspace*{-0.2cm}
\section*{\textsf{Author contributions}}
\vspace*{-0.2cm}
{\small
\noindent
F.H.L.K., P.S. and S.B.P. conceived and designed the experiment. 
S.B.P. and P.S. performed near-field experiments on a system optimized by S.B.P. and N.C.H.H.
Transport experiments were performed by S.B.P. and P.S. on a system built by R.K.K. 
The sample was fabricated by P.S. using hBN crystals provided by K.W. and T.T.
The results were analyzed and interpreted by S.B.P, P.S. and F. H. L. K. using a model developed by D.C, H.H. and B.A.B. 
The manuscript was written by S.B.P., P.S., D.C. and F.H.L.K. with an input from all authors. 
F.H.L.K supervised the work.}

\vspace*{-0.2cm}
\section*{\textsf{Competing Financial Interests}}
\vspace*{-0.2cm}
{\small
\noindent
The authors declare no competing financial interests.
}

\vspace*{-0.2cm}
\section*{\textsf{Data Availability Statement}}
\vspace*{-0.2cm}
{\small
\noindent
The data that support the plots within this paper and other findings of this study are available from the corresponding authors upon reasonable request.
}

\newpage

\end{document}


\title{\Large\textsf{\papertitle}}

\author{Sergi Batlle Porro}
\affiliation{\footnotesize ICFO-Institut de Ciències Fotòniques, The Barcelona Institute of Science and Technology, Av. Carl Friedrich Gauss 3, 08860 Castelldefels (Barcelona),~Spain}

\author{Dumitru C\u{a}lug\u{a}ru}
\affiliation{\footnotesize Department of Physics, Princeton University, Princeton, NJ 08544,~USA}

\author{Haoyu Hu}
\affiliation{\footnotesize Donostia International Physics Center (DIPC),
Paseo Manuel de Lardizàbal. 20018, San Sebastiàn,~Spain}

\author{Roshan Krishna Kumar}
\affiliation{\footnotesize ICFO-Institut de Ciències Fotòniques, The Barcelona Institute of Science and Technology, Av. Carl Friedrich Gauss 3, 08860 Castelldefels (Barcelona),~Spain}

\author{Niels C.H. Hesp}
\affiliation{\footnotesize ICFO-Institut de Ciències Fotòniques, The Barcelona Institute of Science and Technology, Av. Carl Friedrich Gauss 3, 08860 Castelldefels (Barcelona),~Spain}

\author{Kenji Watanabe}
\affiliation{\footnotesize Research Center for Functional Materials, National Institute for Materials Science, 1-1 Namiki, Tsukuba 305-0044,~Japan}

\author{Takashi Taniguchi}
\affiliation{\footnotesize International Center for Materials Nanoarchitectonics, National Institute for Materials Science,  1-1 Namiki, Tsukuba 305-0044,~Japan}

\author{B. Andrei Bernevig}
\affiliation{\footnotesize Department of Physics, Princeton University, Princeton, NJ 08544,~USA}
\affiliation{\footnotesize Donostia International Physics Center (DIPC),
Paseo Manuel de Lardizàbal. 20018, San Sebastiàn,~Spain}
\affiliation{\footnotesize IKERBASQUE, Basque Foundation for Science, 48013 Bilbao,~Spain}

\author{Petr Stepanov}
\email{pstepano@nd.edu}
\affiliation{\footnotesize ICFO-Institut de Ciències Fotòniques, The Barcelona Institute of Science and Technology, Av. Carl Friedrich Gauss 3, 08860 Castelldefels (Barcelona),~Spain}
\affiliation{\footnotesize Department of Physics and Astronomy, University of Notre Dame, Notre Dame, IN 46556,~USA}
\affiliation{\footnotesize Stavropoulos Center for Complex Quantum Matter, University of Notre Dame, Notre Dame, IN 46556,~USA}

\author{Frank H.L. Koppens}
\email{frank.koppens@icfo.eu}
\affiliation{\footnotesize ICFO-Institut de Ciències Fotòniques, The Barcelona Institute of Science and Technology, Av. Carl Friedrich Gauss 3, 08860 Castelldefels (Barcelona),~Spain}
\affiliation{\footnotesize ICREA-Institució Catalana de Recerca i Estudis Avançats, 08010 Barcelona,~Spain}


\maketitle
\def\bibsection{\section*{\textsf{Supplementary references}}} 

\renewcommand{\figurename}{Fig.}
\setcounter{equation}{0}
\setcounter{figure}{0}
\setcounter{table}{0}
\makeatletter
\renewcommand{\theequation}{S\arabic{equation}}
\renewcommand{\thefigure}{S\arabic{figure}}
\renewcommand{\thetable}{\arabic{table}}
\renewcommand{\bibnumfmt}[1]{[#1]}
\renewcommand{\citenumfont}[1]{#1}

\onecolumngrid

\section{Local thermopower measurements}
\label{sec:pv}
\noindent

\begin{figure}[H]
    \centering
    \includegraphics[scale=0.6]{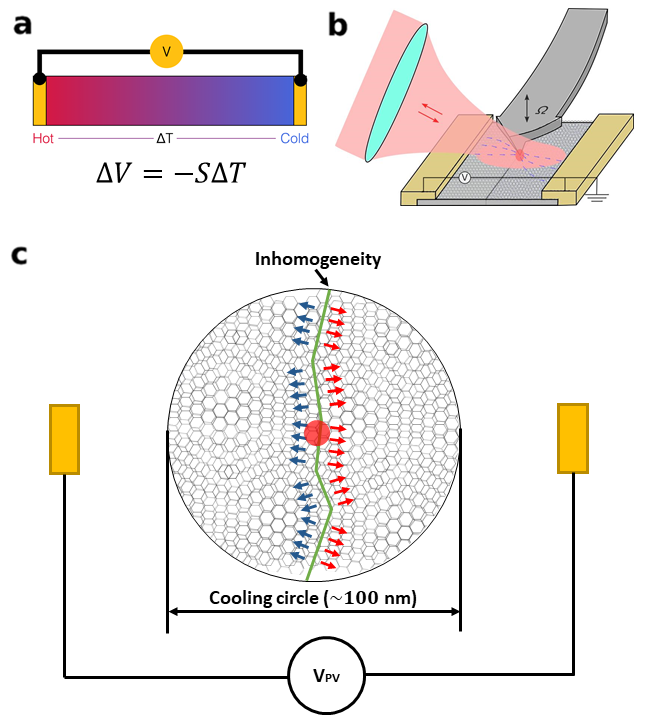}
    \caption{\textbf{Photothermoelectric effect.} \textbf{a}) Standard thermoelectric effect under spatially asymmetric temperature gradient. \textbf{b}) Near-field induced heating creates a cylindrically symmetric temperature gradient around the tip, which in turn produces a voltage gradient in regions with an asymmetric Seebeck coefficient gradient. \textbf{c}) Zoom-in schematics at a junction, where the green line denotes an inhomogeneity (e. g. a domain wall). This results in a gradient of the Seebeck coefficient across the boundary producing a finite thermoelectric voltage.}
    \label{fig:intro}  
\end{figure}

\begin{figure}[H]
    \centering
    \includegraphics[scale=0.75]{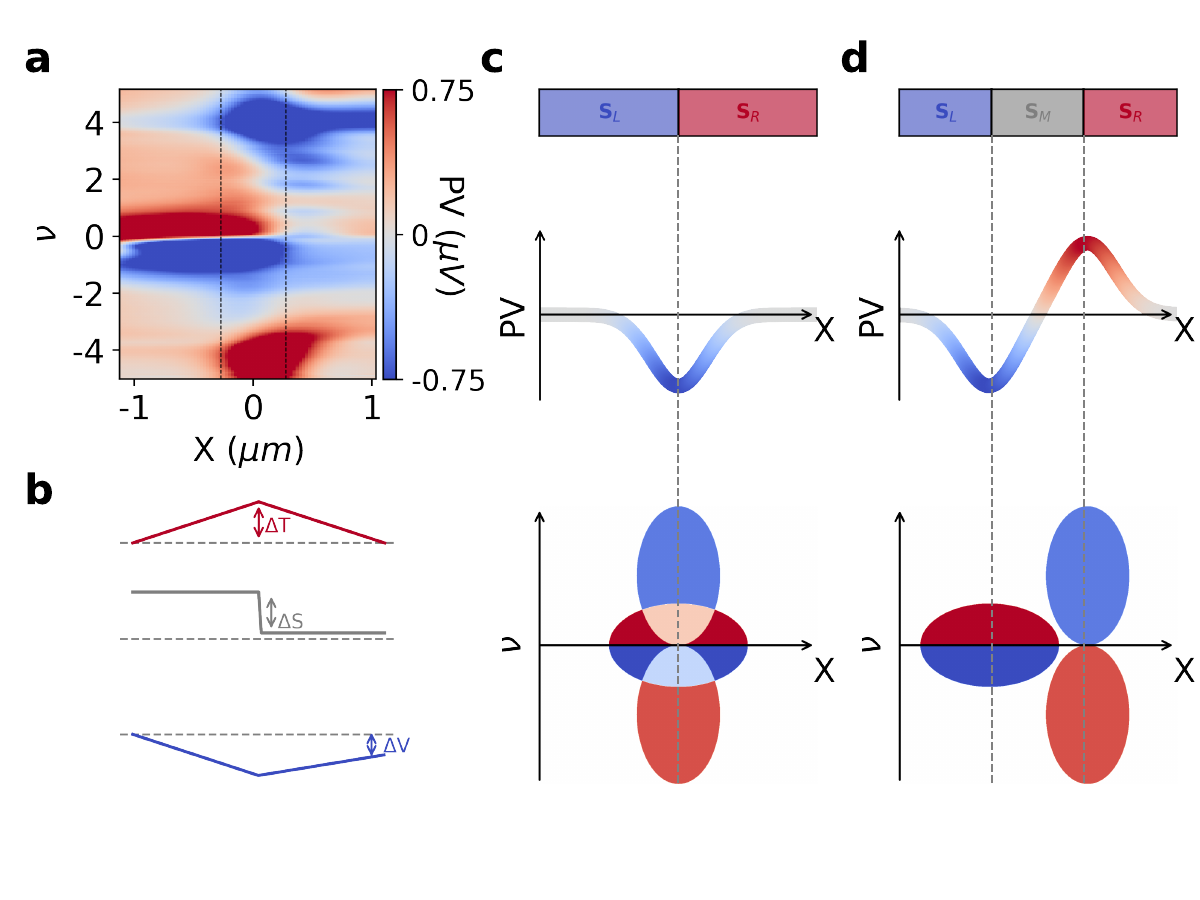}
    \caption{\textbf{Local thermoelectric effect at the interface TG/TSTG.} \textbf{a}) A zoom-in dataset around the TSTG/TG interface shown in the main text Fig. 1b. \textbf{b}) A step-like junction of two different regions with unequal Seebeck coefficients. \textbf{c})(top panel) $PV$ of a single junction. It is expected to be proportional to the difference of the Seebeck coefficients on the right and on the left sides of the interface. (bottom panel) Expected $PV$ vs. $\nu$ and tip position $x$ for the single-junction case. \textbf{d}) (top panel) Double-junction picture. A finite size domain wall separates the two materials with $S_L$ and $S_R$. (bottom panel) Expected $PV$ vs. $\nu$ and tip position $x$ for the double-junction scenario.}
    \label{fig:pv2}  
\end{figure}

In Fig.~\ref{fig:intro} we schematically explain our thermoelectric experiments in the case of a simplified sample geometry (only two voltage probe contacts). Naively, one would expect that our local thermoelectric probe produces equal amounts of charge moving to the left and to the right of the tip (Fig.~\ref{fig:intro}b), hence balancing out the measured $PV$ on both contacts. For an actual device, we anticipate various inhomogeneities within the sample's lattice that would result in a gradient of the Seebeck coefficient (Fig.~\ref{fig:intro}c). We elaborate on this idea in our device further in the Fig.~\ref{fig:pv2}.

A thermoelectric voltage ($\Delta V$) can be generated by the application of a temperature gradient ($\Delta T$) through the Seebeck effect. Fig.~\ref{fig:pv2}a shows a zoom-in map $PV$ vs. tip position $x$ and band filling $\nu$ around the TSTG/TG interface (see Fig. 1a and c in the main text). The generated thermovoltage subsequently depends on the Seebeck coefficient ($S$) and the temperature gradient ($\Delta T$) according to $\Delta V =-S \Delta T$. In our experiment, the near-field enhancement of the electric field generates a hotspot ($\approx$20 nm radius) that moves above the surface (see Fig. 1a in the main text). The heat absorbed via the photothermoelectric effect (PTE) creates a cylindrically symmetric temperature gradient. In an inhomogeneous device (similar to a junction in one dimension), this generates an unequal PTE response on both sides of the junction producing a finite photovoltage drop $\Delta V$. In Fig.~\ref{fig:pv2}b we show a cartoon of such mechanism, where the total voltage is given by the gradients of the Seebeck coefficient ($\Delta S = S_R - S_L$) and the temperature ($\Delta T$): $\Delta V =-\Delta S \Delta T$.

We now focus our attention at the TSTG/TG interface. Since it produces the strongest $PV$ response, it is instructional to describe $PV$ generation close to this interface. Here, we consider two possible scenarios at the junction: 1) Single-junction, i. e. the interface does not have an intermediate low-dimensional channel (Fig.~\ref{fig:pv2}c), and 2) Double-junction with an intermediate channel (Fig.~\ref{fig:pv2}d). In the first case, $PV$ generates has the same values on both sides as shown in the schematic image in Fig.~\ref{fig:pv2}c (upper panel). In this case, expected $PV$ vs. tip position $x$ and band filling $\nu$ is shown schematically in Fig.~\ref{fig:pv2}c (lower panel), where the color stands for the $PV$ amplitude and its sign. Our experimental observations are rather different from this scenario and most consistent with the double-junction picture shown in Fig.~\ref{fig:pv2}d. Here, a region, i. e. a domain wall with a small Seebeck coefficient ($S_M$), separates the two other domains ($S_L$ and $S_R$). Expected $PV$ response is shown in the lower panel and is the most consistent with the experimental data in Fig.~\ref{fig:pv2}a. Vanishing value of the $S_M$ can be explained by the reduced dimensionality of the interface and as a result decreased density of states and transversal velocity of the carriers. This situation creates an optimal opportunity to study thermoelectricity due to a possibility of interpreting the gate- and spatially-resolved photovoltage as the Seebeck coefficient independently on the each side across the interface.

\section{Seebeck coefficient and photovoltage}
\label{sec:mott}
In electron diffusion theory, Mott formula establishes the Seebeck coefficient via the electrical conductivity. It is generally derived from the Kubo thermoelectric transport formula under relaxation time approximation and applies to good metals at low temperatures, where a large density of freely moving carriers is available for transport at the Fermi level. Typically, Mott formula shows a very good agreement in a wide variety of systems including single-layer graphene at room temperature\cite{woessner_near-field_2016, gabor_hot_2011}. Generally, Mott formula takes the following form\cite{jonson_mott_1980}: 

\begin{equation}
S = \frac{\pi^2 k_B^2 T}{3e} \dv{\ln \rho}{V_G} \dv{V_G}{n} \dv{n}{\mu},
\label{eq:Mott_S}
\end{equation}

\noindent
where $S$ is the Seebeck coefficient, $\rho$ is the resistivity, $k_B$ is the Boltzmann constant, $e$ is elementary charge, $T$ is temperature, $V_G$ is the backgate voltage, $n$ is the charge carrier density and $\mu$ is the chemical potential. For equation \ref{eq:Mott_S}, the derivative $\dv{\ln \rho}{\mu}$ was decomposed through chain derivatives to the derivative over the gate voltage $V_G$. The total capacitance $C_T = \frac{1}{e} \dv{n}{V_G}$ can be separated into geometrical and quantum contributions $C_T^{-1}=C_G^{-1}+C_Q^{-1}$, where the quantum capacitance $\frac{C_Q}{e} = \dv{n}{\mu}$. In our case, under the assumption $C_Q \ll C_G$, we get:

\begin{equation}
\dv{V_G}{n} \dv{n}{\mu} = \frac{C_Q}{C_T} = C_Q \left( \frac{1}{C_G} + \frac{1}{C_Q} \right) \approx C_Q/C_Q = 1,
\label{eq:Capacitance}
\end{equation}

\begin{figure}[H]
    \centering
    \includegraphics[scale=0.6]{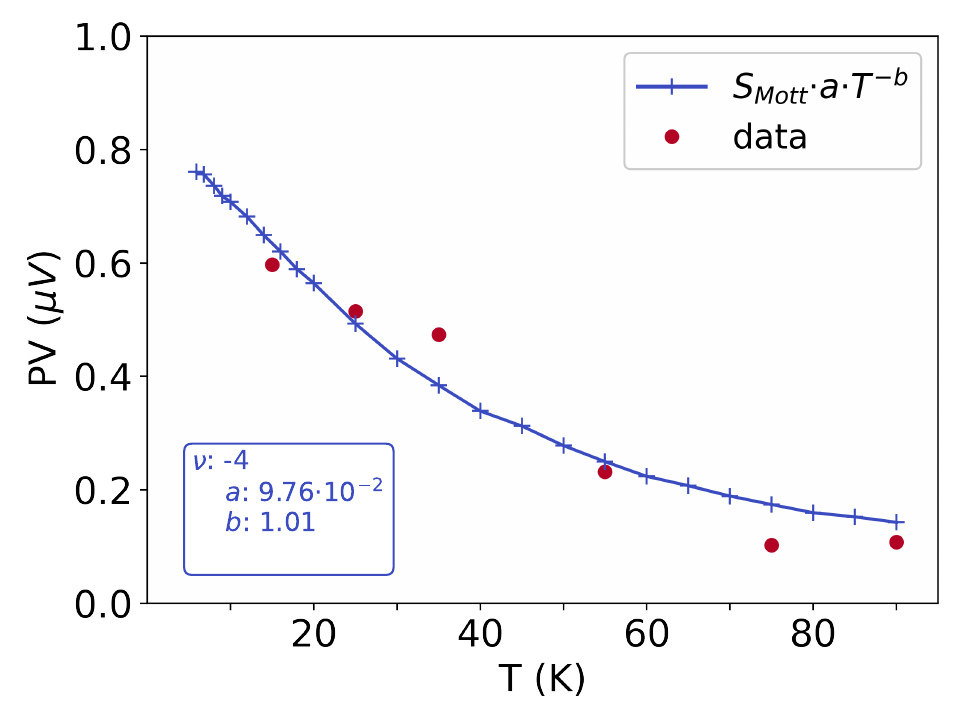}
    \caption{\textbf{Photovoltage temperature dependence} at filling $\nu=-4$ at the same region as Fig. 1 in the main text. Local $PV$ data (red) is fitted (blue) by the Seebeck coefficient calculated by applying equations~\ref{eq:Mott_S},\ref{eq:Capacitance} to the global transport resistivity measurements using the equation~\ref{eq:PVtemp}. A temperature dependence of $\Delta T \propto 1/T$ is obtained.}\label{fig:Tdep}
\end{figure}

The sample temperature $T$ explicitly enters the right-hand side of the equation~\ref{eq:Mott_S}, but the experimental $PV$ decreases with increasing $T$, see figure~\ref{fig:Tdep}. As we show in the SI Section~\ref{sec:pv}, the $PV$ is given by $PV =-S \Delta T$, where $\Delta T$ is the temperature gradient created by the heating from the near-field probe. This heating depends on the infrared light absorption and on the thermal diffusivity of the heterostructure hBN/TSTG/hBN/graphite/SiO2/Si$^{++}$. Both terms can have temperature dependence affected by the intrinsic or extrinsic effects not accounted for in the theoretical calculations. Since the response of our TSTG with the flatbands fully filled  will act like a metal, it can be reasonably well approximated by the Mott formula - as supported qualitatively by figure 1 in the main text -. We use this knowledge to fit the photovoltage to a function with phenomenological shape like in equation~\ref{eq:PVtemp} and extract $\Delta T \propto a T^{-1}$.

\begin{equation}
PV = S \Delta T = -S a T^{-b} ,
\label{eq:PVtemp}
\end{equation}

\section{Effect of the lifetime on the thermopower}
\label{sec:life} 
\noindent
The definition of the Seebeck thermopower given in SI Section~\ref{sec:mott} is insufficient for the theoretical modeling of its microscopic mechanism. Microscopically, the Seebeck coefficient is calculated by computing the correlator functions of the current operators\cite{calugaru_seebeck}. The correlator includes information on the dynamics of the electrons by taking into account both the carriers' dispersion and their lifetime - a major contributor to particle-hole asymmetric $PV$ response in the reported data. This description goes in contrast to the frequently used relaxation approximation \cite{novelli_optical_2020}, which assumes that the lifetime does not depend on the energy of the states. Second-order self-consistent perturbation theory applied to moiré graphene\cite{hu_kondo_2023,song_magic-angle_2022,calugaru_tbg_2023} in the Topological Heavy Fermion paradigm allows us to describe our local thermoelectric measurements beyond Hartree-Fock, so that carriers' lifetimes are naturally accounted for. In this Section, using Kubo formula we derive a phenomenological model that takes into account the finite lifetime of the excitations in addition to their dispersion.

We start with the Kubo formula in its integral form:
\begin{equation}
\sigma S = -e k_B \int \frac{(E-\mu)}{k_B T} D(E) g(E) \left( -\dv{f(E,\mu)}{E} \right) \dd{E},
\label{eq:Kubo}
\end{equation}
\noindent
where $E$ is the energy, $\mu$ is the chemical potential, $g(E)$ is the density of states, $D(E)$ is the electron diffusion coefficient and $f(E,\mu)$ is the Fermi function. In equation~\ref{eq:Kubo}, we define the electrical conductivity $\sigma$ as:

\begin{equation}
\sigma  = e^2 \int D(E) g(E) \left( - \dv{f(E,\mu)}{E} \right) \dd{E}
\label{eq:Conductivity}
\end{equation}

\noindent
Together equations~\ref{eq:Kubo} and~\ref{eq:Conductivity} provide the Seebeck coefficient:

\begin{equation}
S = -\frac{1}{eT} \frac{\int (E-\mu) D(E) g(E) \left(-\dv{f(E,\mu)}{E} \right) \dd{E}}{ \int D(E) g(E) \left( - \dv{f(E,\mu)}{E} \right) \dd{E}}
\label{eq:Seebeck}
\end{equation}

If the electron diffusion coefficient ($D(E)$) is constant, it can be cancelled out from both sides of equation~\ref{eq:Kubo}. However, in the case of heavy-fermion representation of the TSTG's bands, $D(E)$ is unequal across the particle-hole asymmetric minibands. In the most simplified approximation, we can employ $D(E)$ of the free electron gas:

\begin{equation}
D(E) = k_B T \frac{\tau (E)}{m^*}
\end{equation}

\noindent
where $\tau$ is the scattering time and $m^*$ is the effective mass. In a single parabolic band, since $g(E) \propto \theta (E) m^*$ and $-\dv{f(E,\mu)}{E} = \frac{1}{4k_BT} \sech^2 \left( \frac{E-\mu}{2k_BT} \right)$, the resulting Seebeck will be strongly dependent on the scattering time:

\begin{equation}
S = -\frac{1}{eT} \frac{\int(E-\mu) \tau(E) \theta (E) \sech^2\left(\frac{E-\mu}{2k_BT} \right) \dd{E}}{\int \tau(E) \theta (E) \sech^2\left( \frac{E-\mu}{2k_BT} \right) \dd{E}} \propto T \dv{\ln \tau}{\mu}
\label{eq:SeebeckTime}
\end{equation}

\noindent
where in the last equation we applied the Sommerfeld approximation in order to derive a Mott-type Seebeck coefficient, this is only valid for perfect metals. In gapped systems with multiple bands, it is necessary to appropriately take into account each band separately. 

\section{Spectral function around fillings 1 and 2}
\label{sec:band}
\begin{figure}[H]
    \centering
    \includegraphics[scale=0.75]{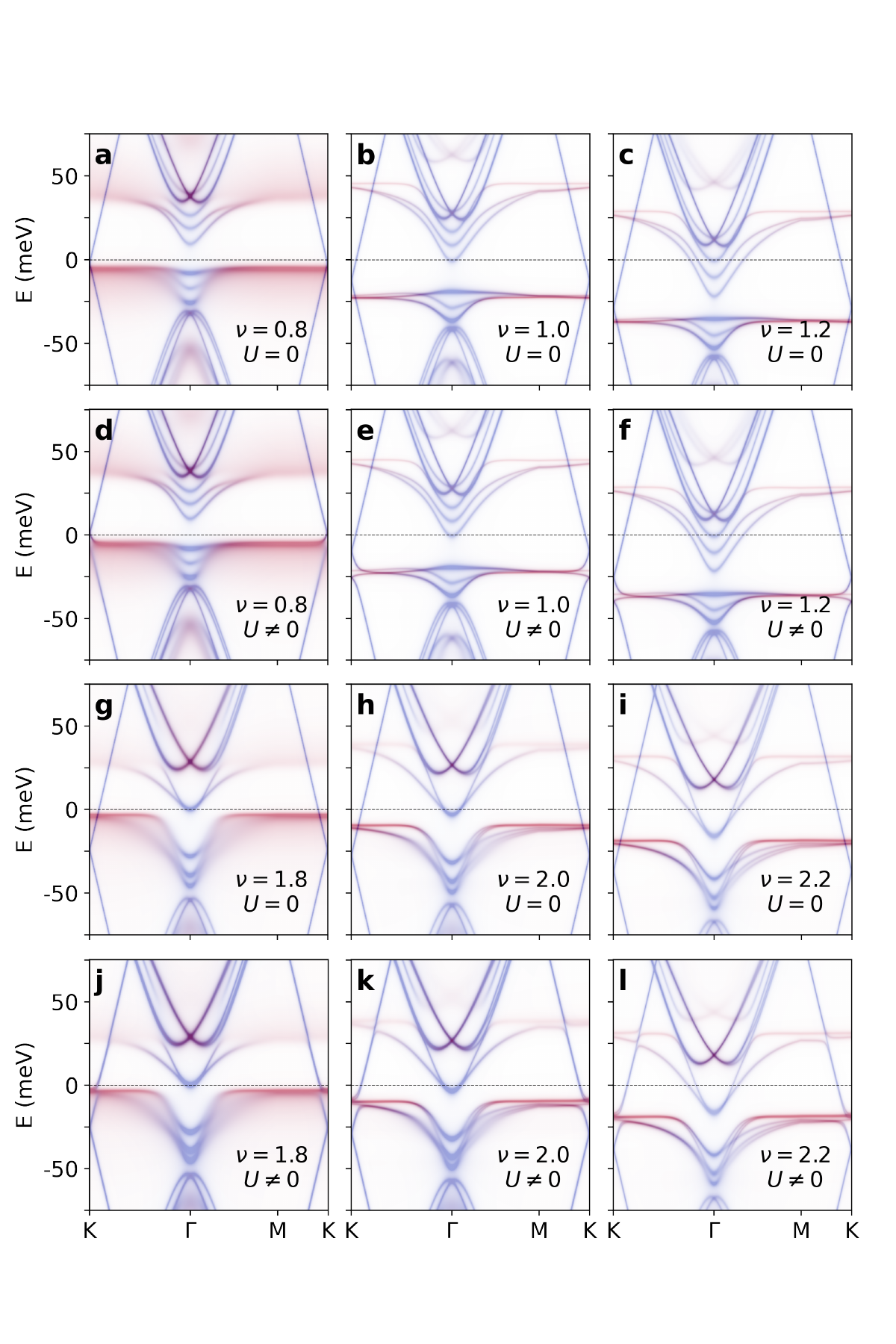}
    \caption{\textbf{Spectral function of TSTG.} Blue color represents \textit{c}-character and red color corresponds to the \textit{f}-character of the charge carriers. \textbf{a-c}) Around $\nu \approx 1$ in the fully Valley Polarized without displacement field $U$ = 0 meV and \textbf{d-f}) with displacement field $U$ = 25 meV, both at $T = 16.8~K$. \textbf{g-i}) Around $\nu \approx 2$ in the KIVC state without displacement field $U$ = 0 meV and \textbf{j-l}) with displacement field $U$ = 25 meV, both at $T = 11.2~K$. For more details, check the article \cite{calugaru_seebeck}.}\label{fig:band2}  
\end{figure}

\noindent
In Fig~\ref{fig:band2} we show the theoretically obtained momentum-resolved spectral functions $A(\omega, \textbf{k})$ for TSTG in the presence and absence of a perpendicularly-applied displacement field ($U$ = 0 (upper panels) and 25 meV (lower panels), respectively). The spectral functions exhibits a heavy, short-lived \textit{f}-character holes and long-lived \textit{c}-character electrons close to the chemical potential set at zero energy. At filling $\nu$ = 2, our experimental data in the main text Fig. 2\textbf{b} shows a dip of the measured $PV$, as well an overall negative offset. This is naturally explained by the theoretically-computed $A (\omega, \textbf{k})$. Around $\nu=2$, the electron excitations have primarly $c$-fermion character. In contrast, being mainly comprised of $f$-fermions, the hole excitations display a highly-incoherent spectral signal (with the spread of the spectral weight being inversely proportional to their lifetime by Heisenberg's principle). As a result, the hole contribution to the Seebeck coefficient is suppressed resulting the negative offset seen in experiments. The Dirac cone electrons ($d$-fermions) have an almost vanishing density of states near the Fermi energy, meaning that their net contribution to thermoelectric transport is negligible\cite{calugaru_seebeck}.

A small application of a displacement field due to the asymmetric gating of the device, does not significantly affect the dispersion of the flat electronic bands. However, it influences the highly dispersive TSTG's Dirac cone. The gap opening of the Dirac cone close to the Fermi level by the displacement field is most significant for transport at filling $\nu$=1.8, where it leads to an enhanced electron contribution to thermelectric transport within a thermally excited window. This added asymmetry shifts the negative $PV$ peak towards lower fillings \cite{calugaru_seebeck}, which is consistent with our experimental observations.

\newpage

\newpage
\section{Extraction of the twist angle and removal of the background signal}
\label{sec:wiggles}
\noindent

\begin{figure}[H]
    \centering
    \includegraphics[scale=0.58]{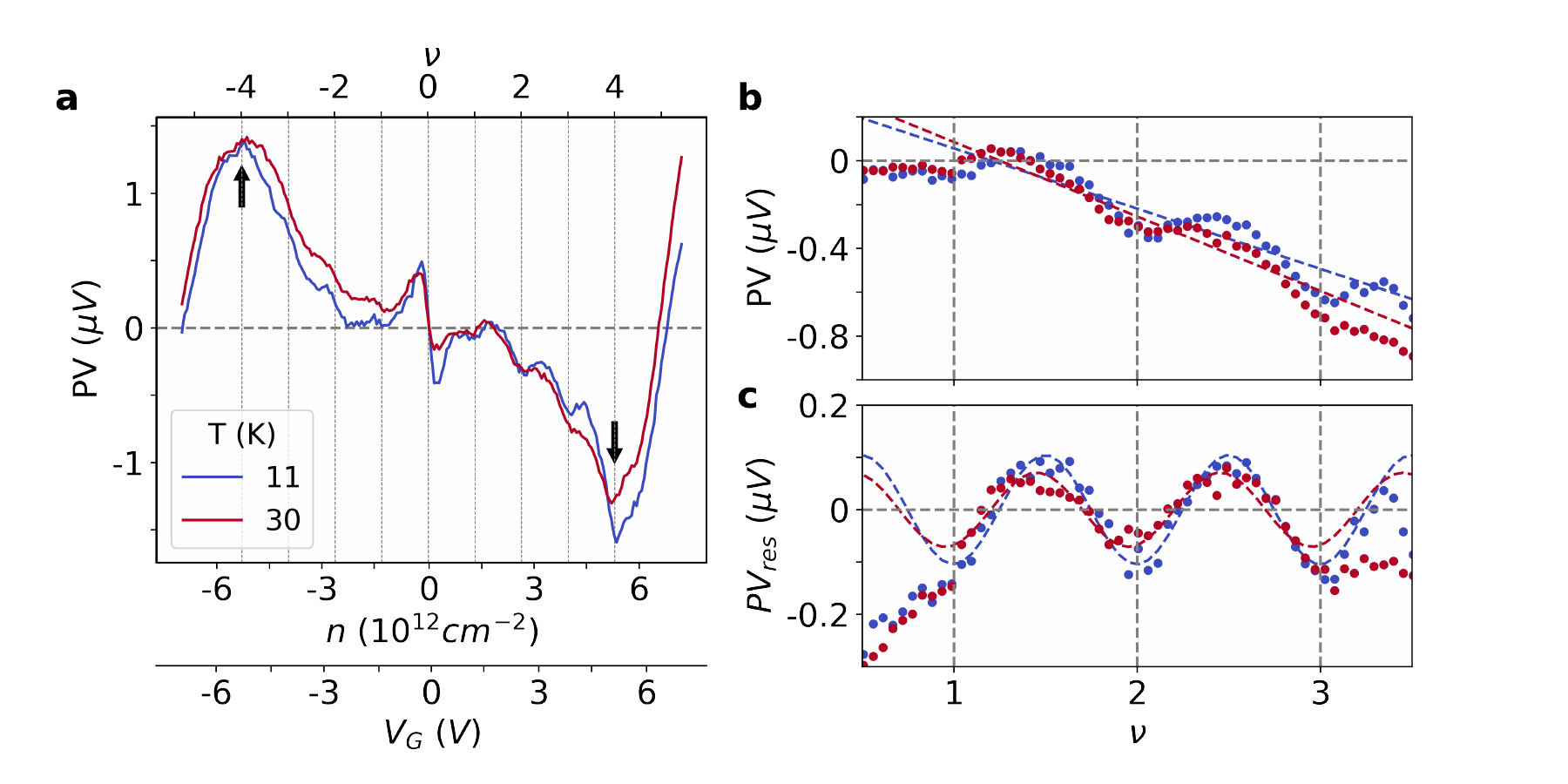}
    \caption{\textbf{Extraction of twist angle and background removal process.} \textbf{a}) $PV$ line traces vs. voltage with the full fillings marked by arrows at $\nu= \pm 4$. For reference, here the $x$-axis has three labels: raw data $V_{bg}$, extracted from Hall resistivity corresponding charge carrier density $n$, and band filling $\nu$. \textbf{b}) $PV$ vs. $\nu$ in the electron-like flatband with the linearly fitted background between $0.5$ and $3.5$. \textbf{c}) Residual of the fitting $PV_{res} = PV-PV_{bkg}$ vs. $\nu$ plotted together with $A\sin{(2 \pi \nu)}+B\cos{(2 \pi \nu)}$ where $A,B$ are extracted using a digital lock-in.}\label{fig:SI_extraction}  
\end{figure}

To obtain the electron filling dependence we scan a line in the sample repeatedly while changing the gate voltage. We extract the photovoltage data using a digital lock-in at the frequency of the tip to the amplified voltage from the sample (See Methods for more details). The phase delay of the electronics is extracted and the photovoltage is projected to real values. In the charge density dependence of twisted graphene samples two strong features emerge: a change of sign at the charge neutrality point (CNP) and two extrema at the full fillings $\nu$ = $\pm$4 (see Fig~\ref{fig:SI_extraction}\textbf{a}). We extract the raw gate voltage values corresponding to each feature. To determine the local twist angle, we convert these values into the charge carrier density. We use a back gate capacitance extracted from the slope of the Hall resistivity vs. gate voltage in the magnetotransport measurements ($C=162.5$~$nF/cm^{-2}$). At the next step, we extract the charge carrier density's local extrema at $\nu$ = $\pm$4 to find an average by using $n_s = \frac{|n_{\nu=-4}|+|n_{\nu=+4|}}{2}$. Finally, we use the relationship between the particle density and the twist angle: $\theta ^2= \sqrt{3} a^2 |n_s|/8$\cite{cao_correlated_2018}, where $a=0.142$~nm is the lattice constant of graphene.

Here, we also describe a process of obtaining the back-ground removed-$PV$ response shown in the main text Fig. 2\textbf{b}. To extract the amplitude of the $PV$ vs. $\nu$ oscillations, we follow an ad-hoc approach. We restrict the data range between $\nu$=$0.5$ and $3.5$ and subtract a constant slope as shown in Fig~\ref{fig:SI_extraction}\textbf{b}. We perform this process independently for the electron and hole-like flat band regions. Once the background is flattened, a discrete digital lock-in is realized by multiplying the obtained curve with $\cos$ and $\sin$ functions of periodicity 1 in the integer filling and numerically integrating the result. 

\begin{align}
A &= \frac{1}{\nu_f-\nu_i} \int_{\nu_i}^{\nu_f} \dd{\nu} \cdot PV_{res}(\nu) \cdot \cos(2 \pi \nu)
\\
B &= \frac{1}{\nu_f-\nu_i} \int_{\nu_i}^{\nu_f} \dd{\nu} \cdot PV_{res}(\nu) \cdot \sin(2 \pi \nu) 
\end{align}

This gives the amplitudes of the Fourier series $A$ and $B$ at with periodicity $\nu$=1 (see fig~\ref{fig:SI_extraction}\textbf{c}):

\begin{equation}
PV_{res}(\nu) \approx A \cdot \cos(2 \pi \nu) + B \cdot \sin(2 \pi \nu) 
\end{equation}

Here we emphasize that this approach remains non-invasive towards the overall position of the $PV$ extrema. Consistent between local $PV$ measurements and electronic transport, the extraction of the $PV$ peaks at $\nu$ = $\pm$4 also guarantees the local twist angle values. To compare consistently between different spatial regions, the final signal is then normalized with the maximum peak at the fully filled band. Thus, the resulting ratio of the amplitude of the oscillations with the full filling $PV$ will not be affected by the strength of the junction or the spatially varying sensitivity of the probe.  

\newpage
\section{Photovoltage data at the interface of TSTG/TG}
\label{sec:junc}
\begin{figure}[H]
    \centering
    \includegraphics[scale=0.8]{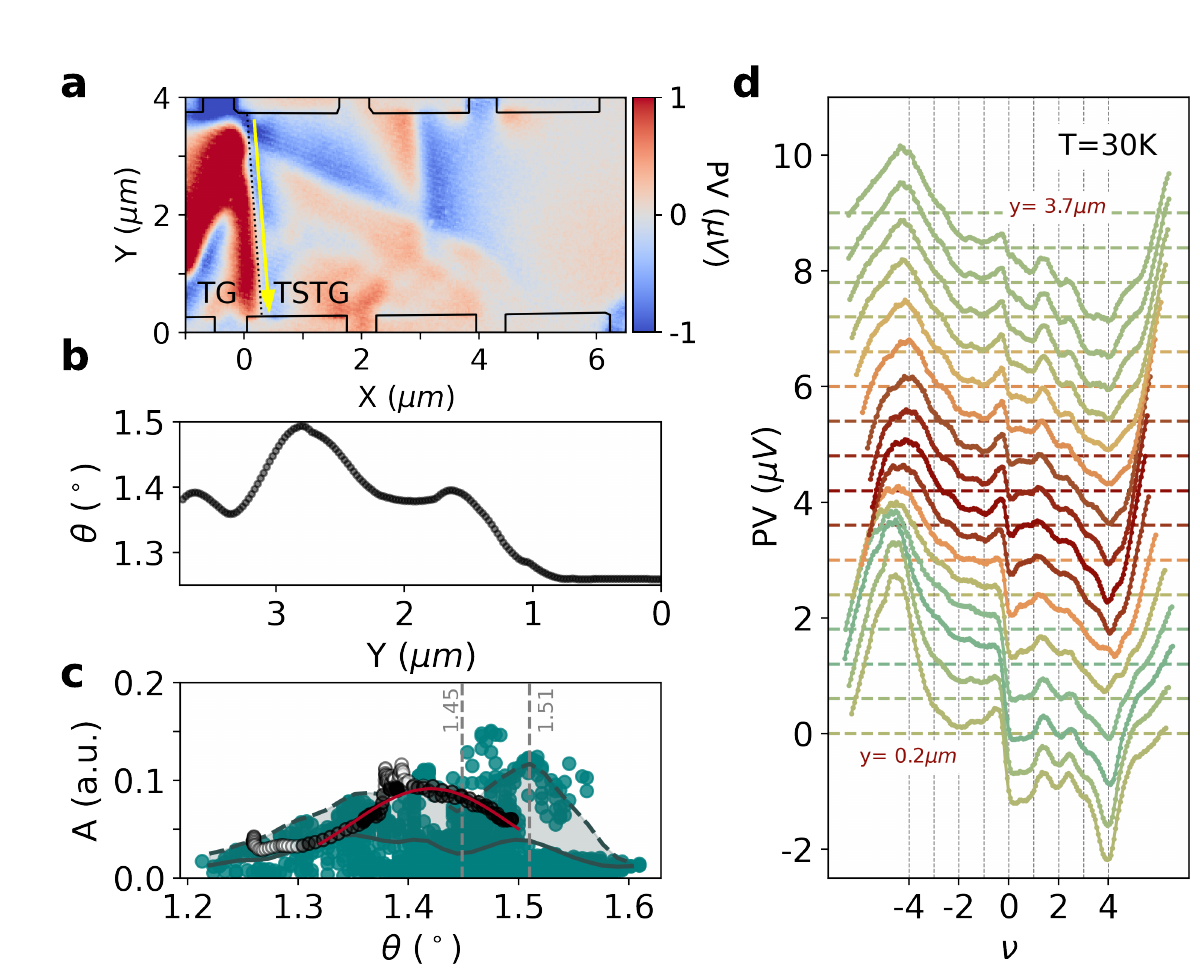}
    \caption{\textbf{Cascade strength at the junction.} \textbf{a}) Linescan location is marked with a yellow arrow in the AFM topography and PV map. \textbf{b} Twist angle extracted vs. position. \textbf{c}) Statistics on the $PV$ magnitude vs. the local twist angle $\theta$ for the electron-like fillings (same a Fig. 5\textbf{b} in main text). Plotted on top, the data from the junction. \textbf{d}) Local $PV$ vs. $\nu$ linetraces taken along the linescan in \textbf{a} and \textbf{b} from top $3.7\mu m$ to bottom $0.2 \mu m$ in equidistant steps.}
    \label{fig:junc2}  
\end{figure}

\noindent
The two materials at the junction provide a unique opportunity to understand thermoelectric transport in moiré materials (see SI Section~\ref{sec:pv}). In the main text Fig. 4\textbf{c}, the $PV$ amplitude is shown for the  distribution of points in the entire TSTG region. In this Section, we also provide data taken very close to the TSTG/TG junction (Fig~\ref{fig:junc2}\textbf{a}). Situated by the proximity to the interface of two materials with very different Seebeck coefficients, $PV$ acquires the largest values. Along the interface, we also extract a wide range of twist angles (Fig~\ref{fig:junc2}\textbf{b}). The twist angle distribution is very similar to the rest of the data in the sample's bulk further confirming the spatial sensitivity of our probe. Any differences in the $PV$ amplitude extracted along the interface and shown in (Fig~\ref{fig:junc2}\textbf{c}) can be explained by a finite subset of sampling in (strain, twist angle) space. Here, we emphasize that the $PV$ sign acquires very strong negative (positive) values on the left(right) side of the interface thus spatially separating thermoelectric probe between the TG and TSTG regions. 

\newpage
\section{Additional local twist angle data}
\label{sec:regions}
\begin{figure}[H]
    \centering
    \includegraphics[scale=0.85]{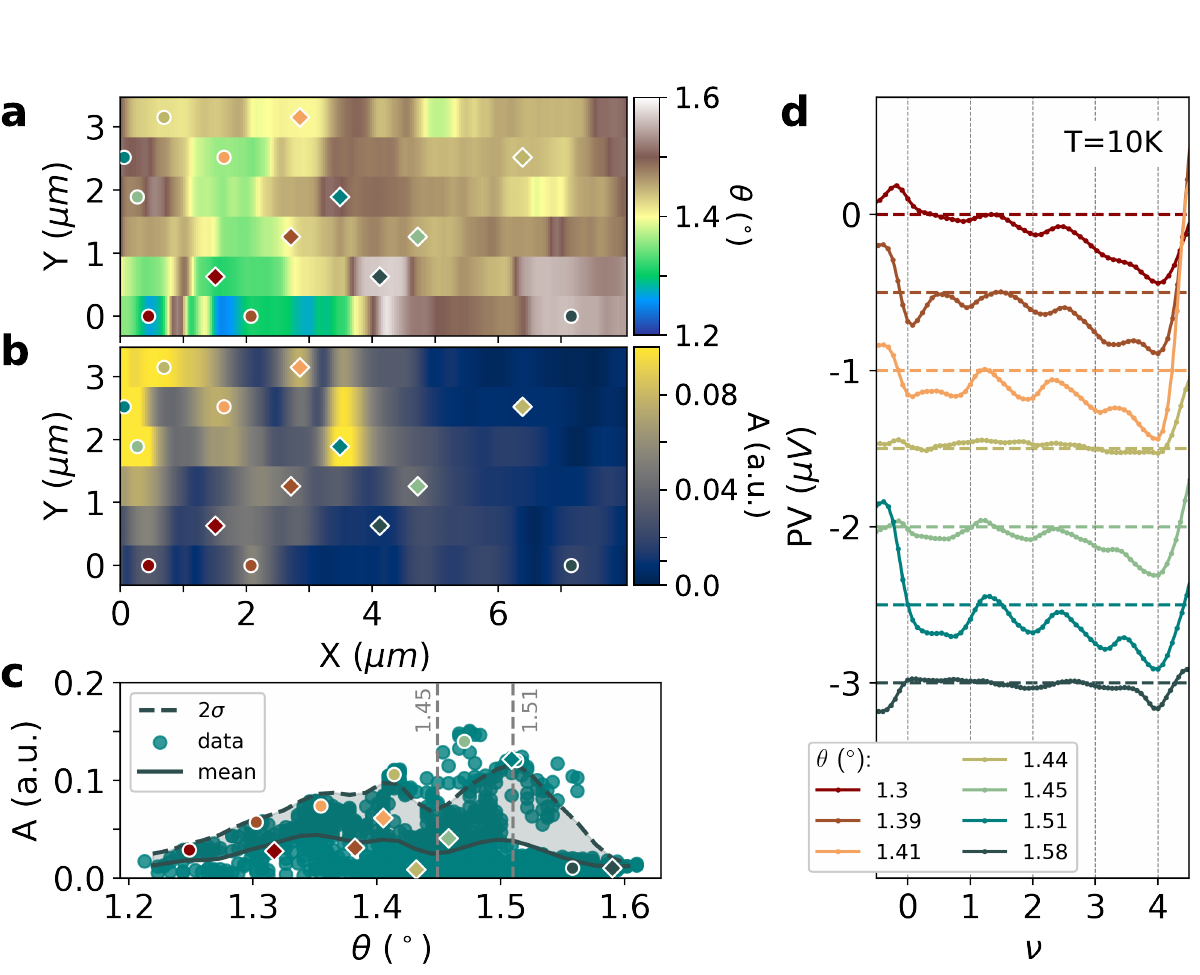}
    \caption{\textbf{Extended twist angle and $PV$ amplitude data.} \textbf{a}) Colorplot of the extracted local twist angle in the TSTG region \textbf{b}) Color plot of the magnitude of the $PV$ vs. $\nu$ oscillations. \textbf{c}) Statistics on the $PV$ magnitude vs. the local twist angle $\theta$ for the electron-like fillings (same a Fig. 5\textbf{b}). \textbf{d}) Local $PV$ vs. $\nu$ linetraces taken at the diamonds shown in \textbf{a},\textbf{b} and \textbf{c}. Data shown in the main text Fig. 4\textbf{c} corresponds to circles.}\label{fig:spatial}  
\end{figure}

Here we present additional data for Fig. 4\textbf{a} in the main text. In Fig.~\ref{fig:spatial} we show additional dataset for tip positions inside the TSTG region. Fig.~\ref{fig:spatial}\textbf{a} and \textbf{b} show local twist angle $\theta$ and $PV$ oscillations amplitude vs. tip position ($x$, $y$), respectively. These two colorplots were used to build Fig. 4\textbf{a}. We add more data from the sample bulk here (diamonds in panels \textbf{a}, \textbf{b}, and \textbf{c}). Color-coded linecuts in Fig.~\ref{fig:spatial}\textbf{d} show the corresponding $PV$ vs. $\nu$ raw data.

Interestingly, some of the data in the sample bulk implies that similar twist angles might have different gate voltage characteristics. Similarly, some of the neighboring regions exhibit similar $PV$ characteristics despite having different twist angles. In TSTG moiré lattice, these regions may have different strain profiles or unequal twist angles between the top and the bottom graphene layers. Our technique allows us to estimate the value of the Seebeck coefficient up to a prior knowledge of the path of the electrical current provided by the Shockley-de-Ramo theorem \cite{song_shockley-ramo_2014}. Thus, our local probe provides the most reliable results at the junction and in between the measurement contacts.

\newpage
\section{Hole-like and electron-like flatbands in TSTG moiré lattice}
\label{sec:ehasym}
\begin{figure}[H]
    \centering
    \includegraphics[scale=0.8]{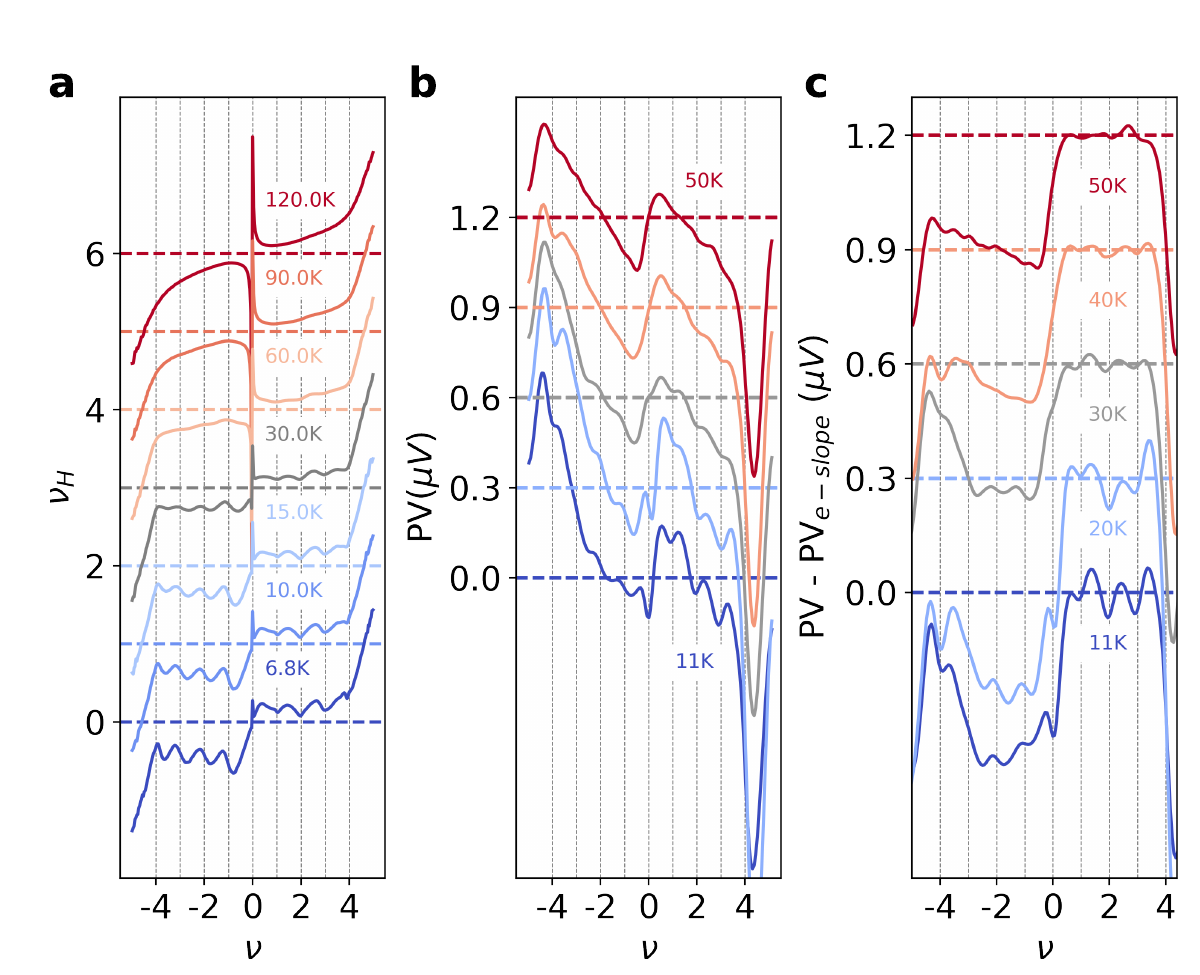}
    \caption{\textbf{Full scale Hall density and PV line traces.} \textbf{a}) Hall density $\nu_H$ vs electron filling $\nu$. \textbf{b}) Local PV traces taken at different temperatures for a fixed AFM tip position. This is also a full density range of the data shown in the main text Fig. 2\textbf{b}. \textbf{c} Same as \textbf{b} but with a linear fit subtracted from the $PV$ in the electron-like flatband.}\label{fig:eh}  
\end{figure}

In this Section we display the data for the hole-like flatband. Fig~\ref{fig:eh}\textbf{a} shows full scale line traces of Hall density $\nu_H$ vs. free charge carrier density $\nu$ for the entire available range from $\nu$ = -5 to $\nu$ = +5. We note a difference between positive and negative fillings. While the Hall density exhibits clear dips at the integer fillings for positive, the negative peaks for negative fillings are slightly shifted towards lower densities than the integer fillings. In addition, there is more Hall density at negative fillings likely pointing at the suppressed effects of the electron localization. This is also consistent with the $PV$ vs $\nu$ shown in Fig~\ref{fig:eh}\textbf{b}, where the $PV$ extrema are a lot less pronounced for $\nu <0$ ($PV$ peaks at $\nu$ = -1 and -2 and disappears at $\nu$ = -3).

For completeness, Fig~\ref{fig:eh}\textbf{b} shows full scale line traces shown in the main text Fig. 2\textbf{b}. Here we note a significant $PV$ slope asymmetry in for negative fillings and electron-like flatbands which becomes even more pronounced at lowest temperatures after a subtraction of a linear fit in the electron region, seeFig~\ref{fig:eh}\textbf{c}. The slope change with temperature both in Hall and $PV$ signals the importance of strong electronic correlations.

\newpage
\section{Magnetoresistance and temperature dependence}
\label{sec:btdep}
\begin{figure}[H]
    \centering
    \includegraphics[scale=0.85]{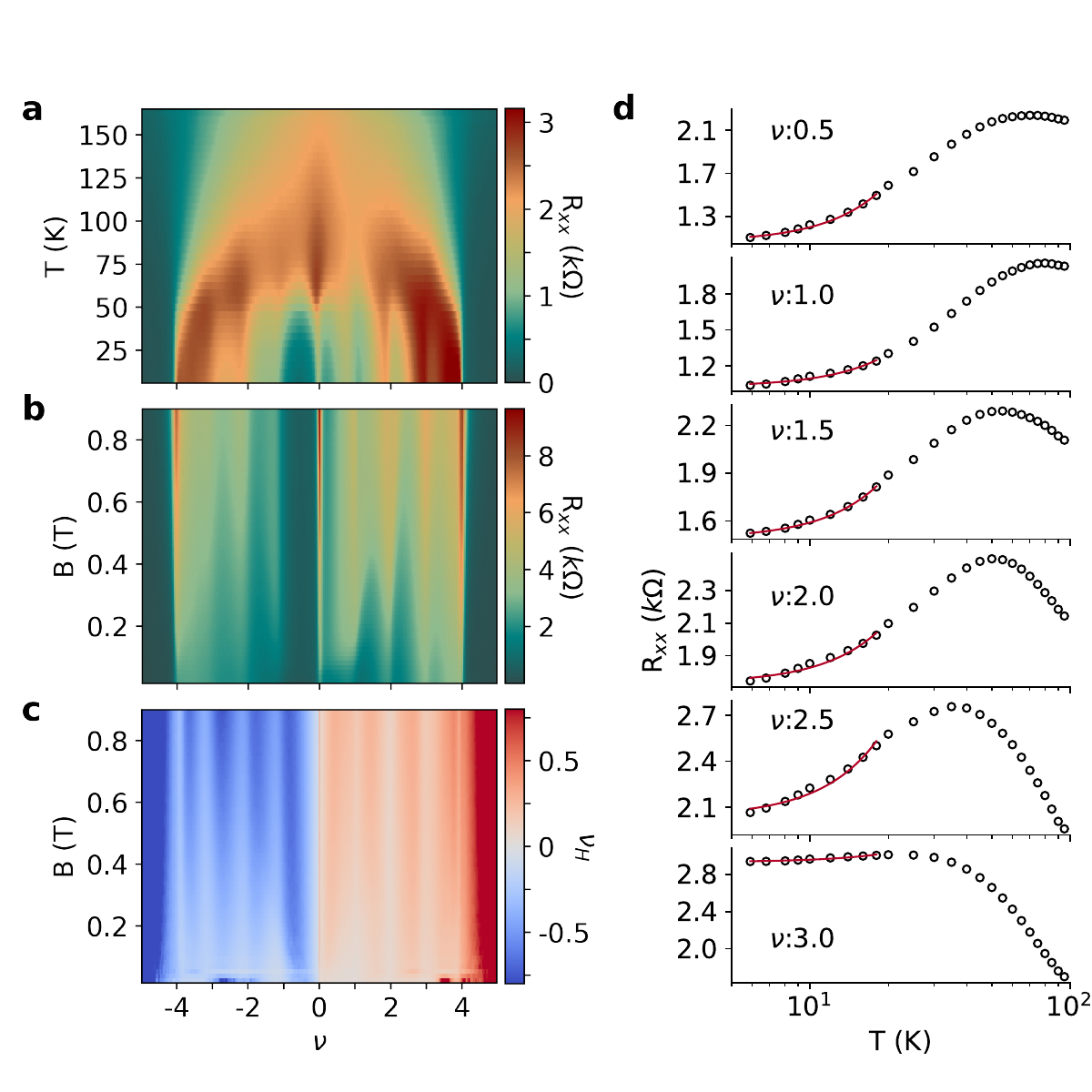}
    \caption{\textbf{Magnetoresistance and temperature dependence of the sample's resistivity.} \textbf{a}) $R_{xx}$ vs. $\nu$ and $T$. \textbf{b}) $R_{xx}$ vs. $\nu$ and $B$ at $T$ = 6.8 K. \textbf{c} $\nu_H$ vs. $\nu$ and $B$ at $T$ = 6.8 K. \textbf{d} Longitudinal resistivity $R_{xx}$ vs. $T$ at different band fillings. Quadratic fit is plotted in red, $R_{xx}(T)=AT^2+B$.}\label{fig:colormaps}  
\end{figure}

In this Section we provide an additional dataset for electronic transport in the TSTG region. Fig~\ref{fig:colormaps}\textbf{a}, \textbf{b}, and \textbf{c} show the longitudinal resistance $R_{xx}$ vs. $\nu$ and $T$ (\textbf{a}), $R_{xx}$ vs. $\nu$ and $B$ (\textbf{b}), and $\nu_H$ vs. $\nu$ and $B$ (\textbf{c}). Fig~\ref{fig:colormaps}\textbf{d} shows a set of linecuts $R_{xx}$ va. $T$ for different band fillings $\nu$. The longitudinal resistance exhibits a maximum between 20-60 K were the slope changes sign, similar to the features observed in other Kondo systems \cite{zhao_gate-tunable_2023}.

\newpage
\subsection{Hall density fraction}
\label{sec:exthall}
\noindent

\begin{figure}[H]
    \centering
    \includegraphics[scale=0.75]{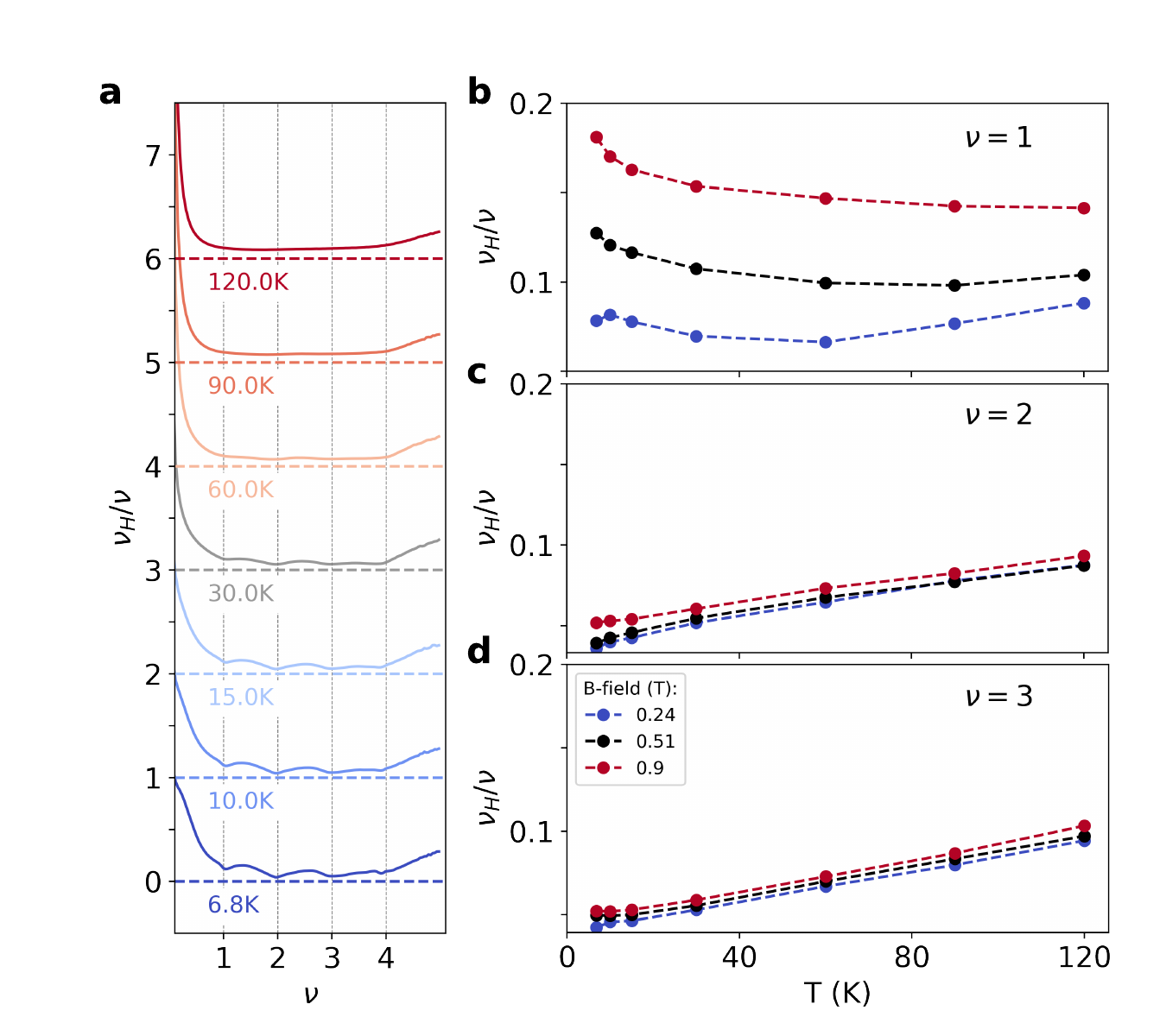}
    \caption{\textbf{Hall density fraction} \textbf{a}) $\nu_H / \nu$ vs $\nu$ at different temperatures $T$ for a fixed magnetic field $B=0.5T$. \textbf{b-d}) $\nu_H / \nu$ vs $T$ at different magnetic fields $B$ for fillings $\nu = 1,2,3$ respectively. Curves are shifted vertically by 1 for clarity.}\label{fig:HallFraction}  
\end{figure}

In this section, we extend the data shown in figure 4 in the main text. In figure~\ref{fig:HallFraction} we observe that the number of electrons contributing to Hall is much smaller than that of the total electrons introduced in the system (below $10 \%$ at low temperatures). This fraction does not grow significantly with doping until above full filling, where each of the carriers added contributes again by one carrier in Hall. In addition, we see a steady increase with temperature at higher fillings and seems not to be there around $\nu = 1$.

\subsection{Magnetoresistance}
\label{sec:magres}
\begin{figure}[H]
    \centering
    \includegraphics[scale=0.8]{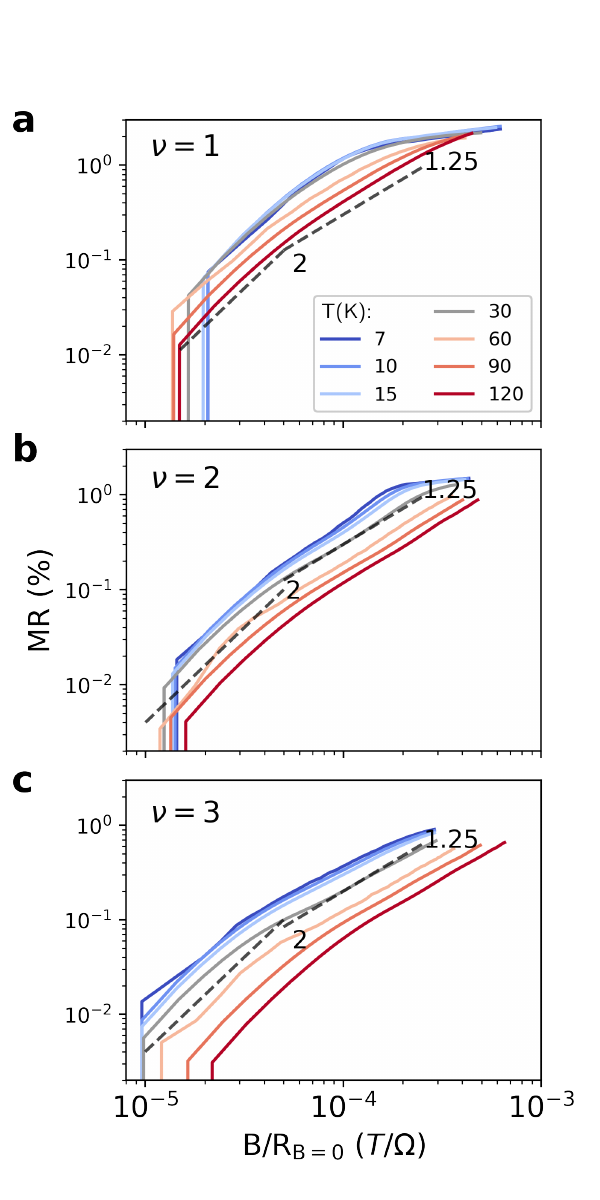}
    \caption{\textbf{Köhler's rule.} Magnetoresistance $R_{xx}$ is plotted as a function of the magnetic field normalized to the scattering (here divided by the resistance at zero field). The data is shown for different temperatures and band fillings \textbf{a}) $\nu = 0$, \textbf{b}) $\nu = 2$ and \textbf{c}) $\nu = 3$. }\label{fig:Kohler}  
\end{figure}

\noindent
To further check our data against the Landau-Fermi liquid theory we explore the quasiparticle character using Köhler's rule shown in Fig~\ref{fig:Kohler} Here, the magnetoresistance $MR = \frac{R(B)-R(0)}{R(0)}$ is plotted against $\frac{B}{R(0)}$ in a typical Köhler's plot for band fillings $\nu$ = +1, +2 and +3. Since the magnetoresistance is $\propto B^2$ at low $B$, this suggests that the heavy fermion quasiparticles have no Plankian behaviour and therefore the system is still away from any possible fluctuations of a quantum critical point \cite{ramires_emulating_2021}. Here, the deviation from order two is more evident at higher fields, where we observe some saturation that might otherwise occur at higher magnetic fields for systems with higher density of states and energy scales.

\newpage

\section{Spatial resolution of our experiment}
\label{sec:resolution}
\noindent
The spatial resolution $L_{\rm res}$ of our experiment is limited by a combination of the AFM tip radius $L_{\rm tip}$ and the inherent spreading $L_{\rm opto}$ attributed to the optoelectronic response. Any features below $L_{\rm res}$ remain unresolved, as they are effectively blurred out. We estimate a resolution of $L_{\rm res} \approx 100$~nm with no temperature dependence similar to other local $PV$ experiments on graphene moiré superlattices\cite{hesp_cryogenic_2023} (see also Fig. 2\textbf{a} in the main text where $PV$ features the same amount of smearing for 10, 50 and 90 K).

\newpage

\noindent
\bibliographystyle{naturemag-sergi}
\bibliography{References.bib}